\documentclass[a4paper,11pt]{article}
\usepackage{jheppub} 

\usepackage{todonotes}
\usepackage{float}
\usepackage{soul}
\usepackage{enumitem}
\usepackage{gensymb}
\newcommand{\tsup}{\textsuperscript}

\newcommand{\trm}{\textrm}
\newcommand{\tbf}{\textbf}
\newcommand{\wh}{\widehat}

\emailAdd{baillysalins@lpsc.in2p3.fr}
\emailAdd{km3net-pc@km3net.de}

\abstract{The existence of an eV-scale sterile neutrino has been proposed to explain several anomalous experimental results obtained over the course of the past 25 years. The first search for such a sterile neutrino conducted with data from KM3NeT/ORCA --- a water Cherenkov neutrino telescope under construction at the bottom of the Mediterranean Sea --- is reported in this paper. GeV-scale atmospheric neutrino oscillations are measured by reconstructing the energy and arrival direction of up-going neutrinos that have traversed the Earth. This study is based on a data sample containing 5828 neutrino candidates collected with 6 detection units ($5\%$ of the complete detector), corresponding to an exposure of 433 kton-years. From the expected effect of an eV-scale sterile neutrino on the first $\nu_\mu \rightarrow \nu_\tau$ standard oscillation maximum, simultaneous constraints are put on the magnitude of the $U_{\mu 4}$ and $U_{\tau 4}$ mixing elements assuming $\Delta m^2_{41} \geq 1$ eV\textsuperscript{2}. The results are compatible with the absence of mixing between active neutrinos and a sterile state, with $|U_{\mu 4}|^2 < 0.138$ and $|U_{\tau 4}|^2 < 0.076$ at a $90\%$ confidence level. Such constraints are compatible with the results reported by other long-baseline experiments, and indicate that with KM3NeT/ORCA it is possible to bring crucial contributions to sterile neutrino searches in the coming years.}

\begin{document}

\title{\boldmath Search for an eV-scale sterile neutrino with the first six detection units of KM3NeT/ORCA}








\author[b,a]{O.~Adriani}
\author[c,be]{A.~Albert}
\author[d]{A.\,R.~Alhebsi}
\author[d]{S.~Alshalloudi}
\author[e]{M.~Alshamsi}
\author[f]{S. Alves Garre}
\author[g]{F.~Ameli}
\author[h]{M.~Andre}
\author[i]{L.~Aphecetche}
\author[j]{M. Ardid}
% ...
\affiliation[a]{Universit{\`a} di Firenze, Dipartimento di Fisica e Astronomia, via Sansone 1, Sesto Fiorentino, 50019 Italy}
\affiliation[b]{INFN, Sezione di Firenze, via Sansone 1, Sesto Fiorentino, 50019 Italy}
\affiliation[c]{Universit{\'e}~de~Strasbourg,~CNRS,~IPHC~UMR~7178,~F-67000~Strasbourg,~France}
\affiliation[d]{Khalifa University of Science and Technology, Department of Physics, PO Box 127788, Abu Dhabi,   United Arab Emirates}
\affiliation[e]{Aix~Marseille~Univ,~CNRS/IN2P3,~CPPM,~Marseille,~France}
\affiliation[f]{IFIC - Instituto de F{\'\i}sica Corpuscular (CSIC - Universitat de Val{\`e}ncia), c/Catedr{\'a}tico Jos{\'e} Beltr{\'a}n, 2, 46980 Paterna, Valencia, Spain}
\affiliation[g]{INFN, Sezione di Roma, Piazzale Aldo Moro, 2 - c/o Dipartimento di Fisica, Edificio, G.Marconi, Roma, 00185 Italy}
\affiliation[h]{Universitat Polit{\`e}cnica de Catalunya, Laboratori d'Aplicacions Bioac{\'u}stiques, Centre Tecnol{\`o}gic de Vilanova i la Geltr{\'u}, Avda. Rambla Exposici{\'o}, s/n, Vilanova i la Geltr{\'u}, 08800 Spain}
\affiliation[i]{Subatech, IMT Atlantique, IN2P3-CNRS, Nantes Universit{\'e}, 4 rue Alfred Kastler - La Chantrerie, Nantes, BP 20722 44307 France}
\affiliation[j]{Universitat Polit{\`e}cnica de Val{\`e}ncia, Instituto de Investigaci{\'o}n para la Gesti{\'o}n Integrada de las Zonas Costeras, C/ Paranimf, 1, Gandia, 46730 Spain}
% ...
\affiliation[be]{Universit{\'e} de Haute Alsace, rue des Fr{\`e}res Lumi{\`e}re, 68093 Mulhouse Cedex, France}
% ...

\maketitle
\flushbottom

\section{Introduction \label{s:intro}}

Since their first observation in 1998 \cite{super-kamiokandecollaborationEvidenceOscillationAtmospheric1998, ambrosioMeasurementAtmosphericNeutrinoinduced1998}, neutrino oscillations have been modeled through the three-flavour Pontecorvo-Maki-Nakagawa-Sakata (PMNS) paradigm \cite{dentonSnowmassNeutrinoFrontier2022}. The model describes the mixing between the neutrino flavour eigenstates ($\nu_e$, $\nu_\mu$, $\nu_\tau$) and the neutrino mass eigenstates ($\nu_1$, $\nu_2$, $\nu_3$) of masses ($m_1$, $m_2$, $m_3$). In this formalism, the PMNS matrix is a $3$-by-$3$ rotation matrix parametrised by the three mixing angles $\theta_{12}$, $\theta_{13}$ and $\theta_{23}$ and the Dirac CP-violating phase $\delta$. The mixing angles determine the extent of mixing between the neutrino states, while the CP-violating phase breaks the symmetry of oscillation probabilities for neutrinos and antineutrinos. The frequencies of the observed oscillations are determined by the values of two independent squared-mass differences, for example $\Delta m^2_{31}$ and $\Delta m^2_{21}$. Over the past 25 years, experiments using various neutrino sources have significantly improved the precision in measuring these six oscillation parameters, with uncertainties currently down to the percent level  \cite{estebanFateHintsUpdated2020}. However, the $\delta$ phase is still largely unconstrained by experiments. The $\theta_{23}$ octant (whether it is lower than, equal to or higher than $\pi/4$) is also undetermined. Finally, the sign of $\Delta m^2_{31}$, determining the neutrino mass ordering (NMO) which is either normal ($m_1 < m_2 < m_3$, NO) or inverted ($m_3 < m_1 < m_2$, IO), is also unknown. Current and upcoming large-scale neutrino oscillation experiments will pursue these questions over the next decade to deepen our understanding of neutrino properties.\\

Despite the success of the three-flavour PMNS paradigm at describing the neutrino oscillation data, several experimental results obtained at a relatively short baseline (SBL) remain largely unexplained \cite{aceroWhitePaperLight2023}. All these anomalies, despite originating from experiments with different neutrino sources, detection technologies, oscillation channels, baselines and neutrino energies, can be individually interpreted as oscillations driven by a squared-mass difference $\Delta m^2 \sim \mathcal{O}(1)$ eV$^2$. This value of $\Delta m^2$ is several orders of magnitude larger than the measured values of $\Delta m^2_{31}$ and $\Delta m^2_{21}$. Thus, to explain the SBL anomalies in terms of oscillations, a simple extension of the standard model is to assume the existence of a fourth, eV-scale, neutrino state. However, cosmological constraints exist on the effective number of relativistic neutrino species \cite{lesgourguesMassiveNeutrinosCosmology2006, particledatagroupReviewParticlePhysics2024}. In addition, the LEP experiments obtained compelling evidence that only three flavours of light neutrinos that couple to the weak interaction --- known as active neutrinos --- exist \cite{PrecisionElectroweakMeasurements2006}.\footnote{In that context, light means with a mass lower than half of the mass of the Z boson.} Thus, an additional eV-scale neutrino would have to be insensitive to the weak interaction, or sterile. In the following, this model will be referred to as the 3$+$1 model.\\

The presence of an eV-scale sterile neutrino would also alter the oscillation probabilities of neutrinos detected in experiments with longer baselines such as atmospheric neutrino telescopes. The IceCube collaboration recently reported updated results of searches for an eV-scale sterile neutrino using their complete detector \cite{icecubecollaborationSearchEVscaleSterile2024a, abbasiExplorationMassSplitting2024a} as well as using the denser DeepCore array \cite{icecubecollaborationSearchLightSterile2024}. In the present work, the first search for an eV-scale sterile neutrino performed with data from the KM3NeT/ORCA detector is presented. After a review of the effect of an eV-scale sterile neutrino on the oscillation channels of interest in section \ref{s:oscprob}, the KM3NeT/ORCA water Cherenkov neutrino telescope is presented in section \ref{s:det}. The dataset used in this study was collected with an early partial configuration of the detector with six detection units, called ORCA6 in the following, operating from January 2020 to November 2021. The data sample as well as the event selection strategy is presented in section \ref{s:samplE_sel}. The model and statistical methods used to perform this oscillation analysis are presented in section \ref{s:methods}. Finally, the results are summarised and discussed in sections \ref{s:res} and \ref{s:ccl} respectively.

\section{Effect of an eV-scale sterile neutrino on the oscillation probabilities \label{s:oscprob}}

Oscillations in the presence of a single sterile neutrino can be modeled by extending the standard three-flavour formalism to include a fourth mass eigenstate $m_4$. This extension introduces one additional squared-mass difference, usually denoted by $\Delta m_{41}^2$. The PMNS matrix becomes a $4\times 4$ unitary matrix $U_{\alpha i}$ with $\alpha=e, \mu, \tau, s$ and $i=1,2,3,4$, which can be parametrised such that:
\begin{equation}
U = R_{34}\tilde{R}_{24}\tilde{R}_{14}R_{23}\tilde{R}_{13}R_{12}\,,
\end{equation}
where $R_{jk}$ is a rotation matrix in the $j$-$k$ plane and, similarly, $\tilde{R}_{jk}$ is a unitary rotation matrix with an added complex phase.
In addition to the three standard mixing angles and the Dirac CP violation phase, three active-sterile mixing angles $\theta_{i4}$, $i=1,2,3$, and two additional Dirac CP violation phases $\delta_{i4}$, $i=1,2$ are usually introduced to parametrise the extended matrix. The active-sterile mixing angles are related to the active-sterile mixing elements by:
\begin{equation}
U_{e4} = \sin \theta_{14} e^{-i\delta_{14}}
\end{equation}
\begin{equation} \label{e:mu}
U_{\mu 4} = \cos \theta_{14} \sin \theta_{24} e^{-i\delta_{24}}
\end{equation}
\begin{equation} \label{e:tau}
U_{\tau 4} = \cos \theta_{14} \cos \theta_{24} \sin \theta_{34}\,.
\end{equation}

In addition to the intrinsic vacuum effect of the added mass eigenstate on the oscillation probabilities, matter effects are critically important in the 3+1 model. When active neutrinos propagate through a medium, they experience the weak potential induced by the nucleons and electrons within the matter. This leads to coherent forward scattering \cite{wolfensteinNeutrinoOscillationsMatter1978}, which can occur for all active flavours via neutral current (NC) interactions with neutrons, protons and electrons, and for $\nu_e$ or $\bar{\nu}_e$ via charged current (CC) interactions with electrons. For the CC channel, the corresponding effective potential is \cite{giuntiFundamentalsNeutrinoPhysics2007}: 

\begin{equation}
	V_\trm{CC} = \pm \sqrt{2} G_F N_e\,,
\end{equation}
where $G_F$ is the Fermi constant, $N_e$ the electron density of the medium and $V_\trm{CC}$ is positive for $\nu_e$, negative for $\bar{\nu}_e$. Neutral current interactions with neutrons, protons and electrons are equally possible for $\nu_e$, $\nu_\mu$ and $\nu_\tau$. The corresponding potential is:
\begin{equation}
	V_\trm{NC} = V_\trm{NC}^{n} + V_\trm{NC}^{p} + V_\trm{NC}^{e^-}
\end{equation}
where, due to equal densities of protons and electrons in ordinary matter and to their opposite charge, the potentials $V_\trm{NC}^{p}$ and $V_\trm{NC}^{e^-}$ cancel out. Only the neutron-induced potential remains, yielding:
\begin{equation}
	V_\trm{NC} = \mp\frac{1}{2} \sqrt{2} G_F N_n
	\label{e:V_NC}
\end{equation}
where $N_n$ is the neutron density of the medium and $V_\trm{NC}$ is negative for $\nu$, positive for $\bar{\nu}$.

In the standard three-flavour case, matter effects arise as only electron (anti)neutrinos are affected by $V_\trm{CC}$, impacting their propagation and altering the oscillation patterns with respect to the vacuum case. However, all three flavours experience the same NC potential from neutrons, so no flavour-dependent coherent forward scattering is introduced by $V_\trm{NC}$. Conversely, in the 3$+$1 case, sterile neutrinos are not affected by $V_\trm{NC}$, with the neutrino propagation being described by the following Hamiltonian in the flavour basis:

\begin{equation}
	H_\textrm{fl} = U H_0 U^\dagger + V_{4\nu}
\end{equation}
where $U$ is the 3$+$1 PMNS matrix in vacuum, $H_0=\frac{1}{2E}\textrm{diag}(0,\Delta m^2_{21},$ $\Delta m^2_{31}, \Delta m^2_{41})$ is the Hamiltonian in vacuum in the mass basis and $V_{4\nu}=\textrm{diag}(V_\trm{CC},0,0,$ $-V_\trm{NC})$, where $V_\trm{NC}$ has been subtracted from the diagonal.\\

\begin{figure}[h!]
  \centering
  	\includegraphics[width=0.49\linewidth, trim = {1.5cm 0.5cm 0.5cm 0.5cm }, clip=True]{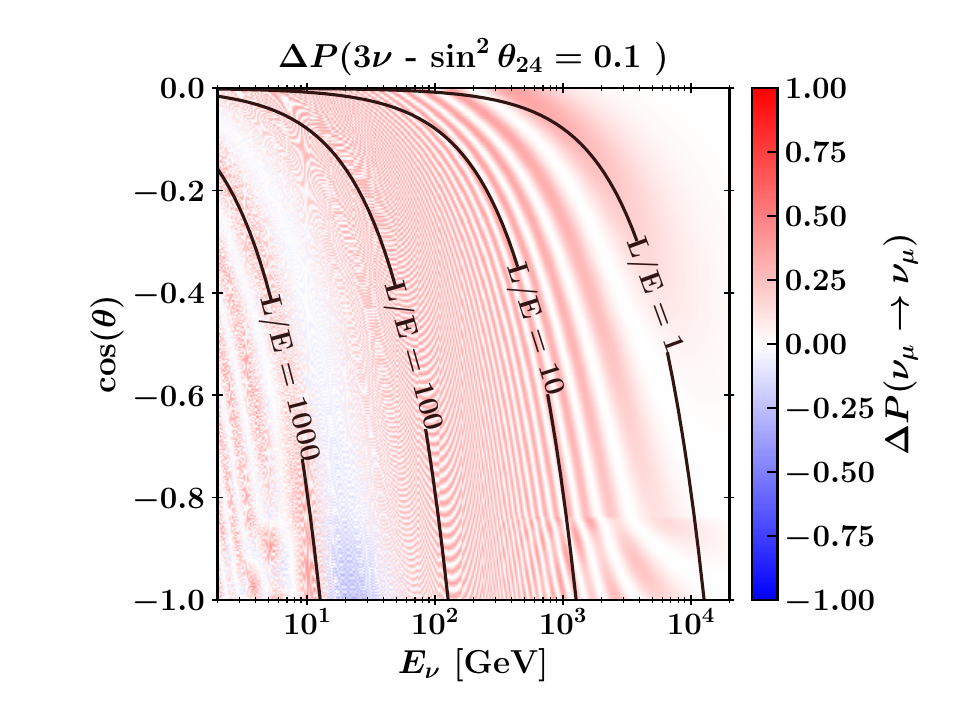}
  	\includegraphics[width=0.49\linewidth, trim = {1.5cm 0.5cm 0.5cm 0.5cm }, clip=True]{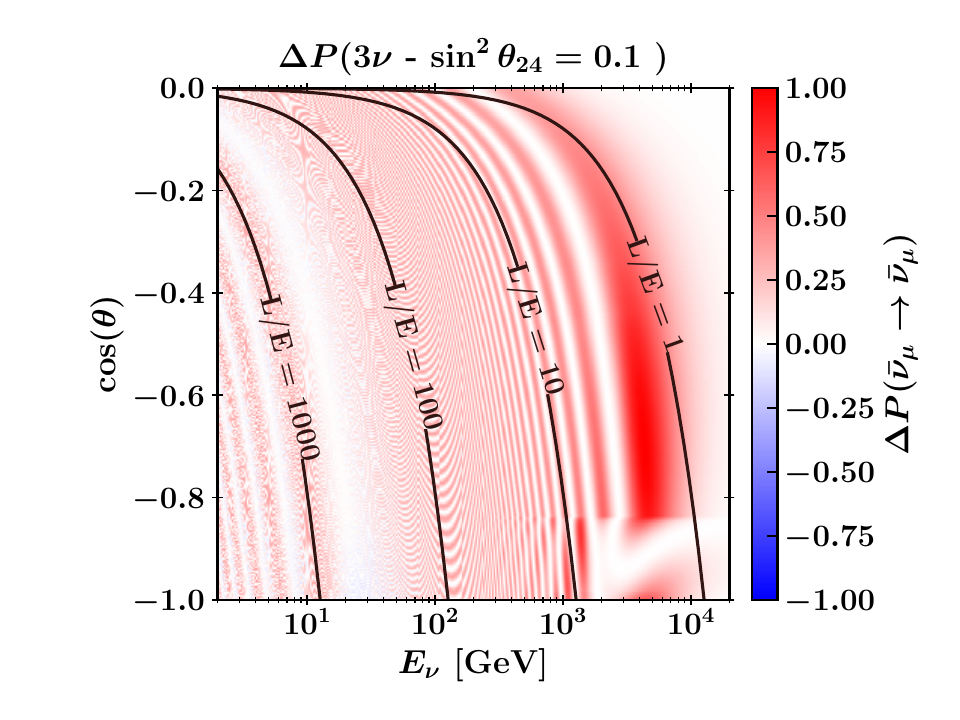}
  \caption{Difference in muon neutrino (left) and antineutrino (right) survival probability between the standard three-flavour model and a 3$+$1 model with $\Delta m^2_{41} = 1$ eV\tsup{2} and $\sin^2 \theta_{24} = 0.1$ for up-going neutrinos traveling through the Earth. Normal ordering is assumed and the standard oscillation parameter values are fixed to the NuFit 5.2 ones \cite{estebanFateHintsUpdated2020}. The other sterile parameters are set to $0$. The L/E lines are shown in units of km/GeV. The oscillograms presented in this article are all computed with OscProb \cite{joao_coelho_2023_10104847} using the Preliminary Reference Earth Model \cite{dziewonskiPreliminaryReferenceEarth1981}.}
  \label{f:sterile_1eV_2-dimensional_numu}
\end{figure}

When accounting for matter effects in the 3$+$1 case, the oscillation probabilities become very complex and difficult to express analytically. Some approximations can be made for specific regimes of $\Delta m^2_{41}$ and neutrino energy, as shown in \cite{aielloSensitivityLightSterile2021}. Preliminary studies \cite{bailly-salinsAtmosphericMuonStudies2024} showed that ORCA6 has no sensitivity to the $\theta_{14}$ mixing angle, as it mainly affects the $\nu_e$ disappearance channel, where the detector's sensitivity is limited. Thus, $\theta_{14}=0$ (and consequently $\delta_{14}=0$) is considered in the following. An analytical formula for $P(\nu_\mu \rightarrow \nu_\mu)$ (not used in this work) has been derived under the assumption $\theta_{14}=\theta_{13}=\theta_{12}=0$ in \cite{super-kamiokandecollaborationLimitsSterileNeutrino2015} and extended in the appendix of \cite{albertMeasuringAtmosphericNeutrino2019} to include the effect of $\delta_{24}$. The preliminary studies in \cite{bailly-salinsSensitivityKM3NeTORCA62024} also showed that ORCA6 has poor sensitivity to very light sterile neutrinos with $\Delta m_{41}^2 < 10^{-2}$ eV\tsup{2} and that its sensitivity to both $\theta_{24}$ and $\theta_{34}$ is mostly independent of $\Delta m_{41}^2 > 10^{-1}$ eV\tsup{2}, as the fast oscillations induced by increasing values of $\Delta m_{41}^2$ are smeared out due to detector effects. Thus, for simplicity, $\Delta m_{41}^2 = 1$ eV\tsup{2} is considered in the following, and the results are assumed to be valid for $\Delta m_{41}^2 \geq 1$ eV\tsup{2}. The difference in $\nu_\mu$ survival probability between the three-flavour case and the 3$+$1 case with $\Delta m^2_{41} = 1$ eV\textsuperscript{2} and $\sin^2 \theta_{24} = 0.1$ (all other sterile parameters being set to 0) is shown in figure \ref{f:sterile_1eV_2-dimensional_numu}, where normal ordering is assumed. The probabilities are shown for atmospheric neutrinos (left) and antineutrinos (right) crossing the Earth, as a function of the neutrino energy $E_\nu$ and the cosine of its zenith angle $\theta$. The zenith angle is defined as the angle between the neutrino direction and the upwards vertical at the detector position, and is directly related to the neutrino propagation length inside the Earth, $L$.\footnote{$\cos \theta = -1$ corresponds to vertical up-going neutrinos and $\cos \theta = 0$ to horizontal neutrinos.} The effects of the fast $\nu_\mu \rightarrow \nu_s$ oscillations can be seen in both cases, lowering the $\nu_\mu$ survival probability on average in the whole energy and $\cos \theta$ range. Two striking features appear in the eV sterile case: on the left,  the reduced $\nu_\mu$ disappearance at around $25$ GeV for vertical neutrinos; and on the right, the enhanced $\bar{\nu}_\mu$ disappearance at TeV energies. Although they happen at very different energies, both are matter-induced effects.\\ 

The TeV $\bar{\nu}_\mu$ disappearance resonance was established more than twenty years ago as a distinctive signature of an eV-scale sterile neutrino in atmospheric neutrino oscillations  \cite{nunokawaProbingLSNDMass2003}. As seen in figure \ref{f:sterile_1eV_2-dimensional_numu}, the effect is more prominent at $E_\nu \simeq 5$ TeV for mantle-only crossing trajectories ($\cos \theta > -0.838$). It can be interpreted as the consequence of the Mikheyev-Smirnov-Wolfenstein effect \cite{wolfensteinNeutrinoOscillationsMatter1978, ermilovav.k.BuildupNeutrinoOscillations1986, mikheyevResonantNeutrinoOscillations1989} in the Earth's mantle. The $\bar{\nu}_\mu$ disappearance peak corresponds to a conversion into an almost pure sterile state \cite{bailly-salinsAtmosphericMuonStudies2024}. This feature has been extensively used in analyses of IceCube data to constrain the $\theta_{24}$ mixing angle as a function of $\Delta  m_{41}^2$ \cite{razzaqueSearchingSterileNeutrinos2011, razzaqueSearchesSterileNeutrinos2012, esmailiConstrainingSterileNeutrinos2012, aartsenSearchingEVscaleSterile2020, icecubecollaborationSearchEVscaleSterile2024a}. For KM3NeT/ORCA, the statistics of TeV-scale neutrinos are limited because of its comparatively smaller detector size.\\

\begin{figure}[h!]
  \centering
  \includegraphics[width=0.49\textwidth]{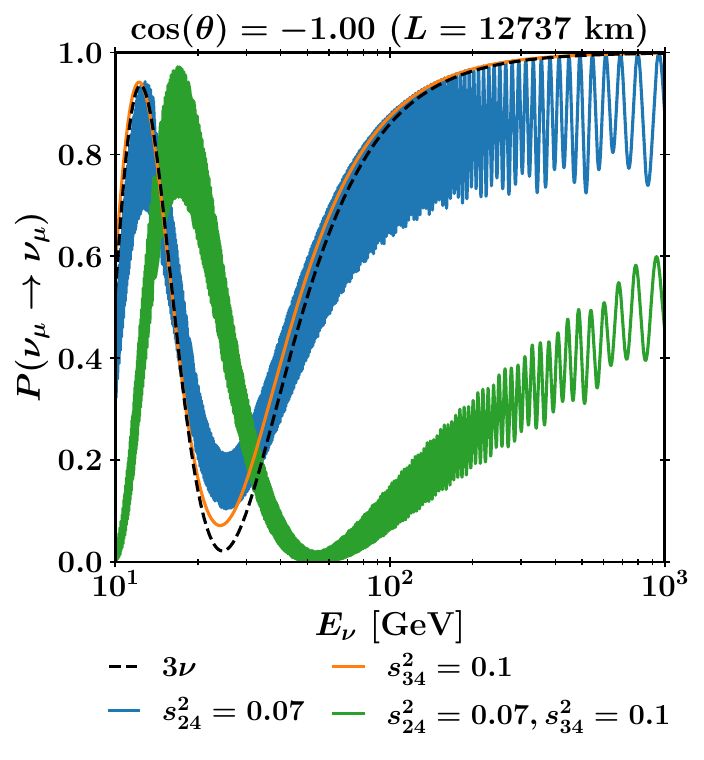}
  \includegraphics[width=0.49\textwidth]{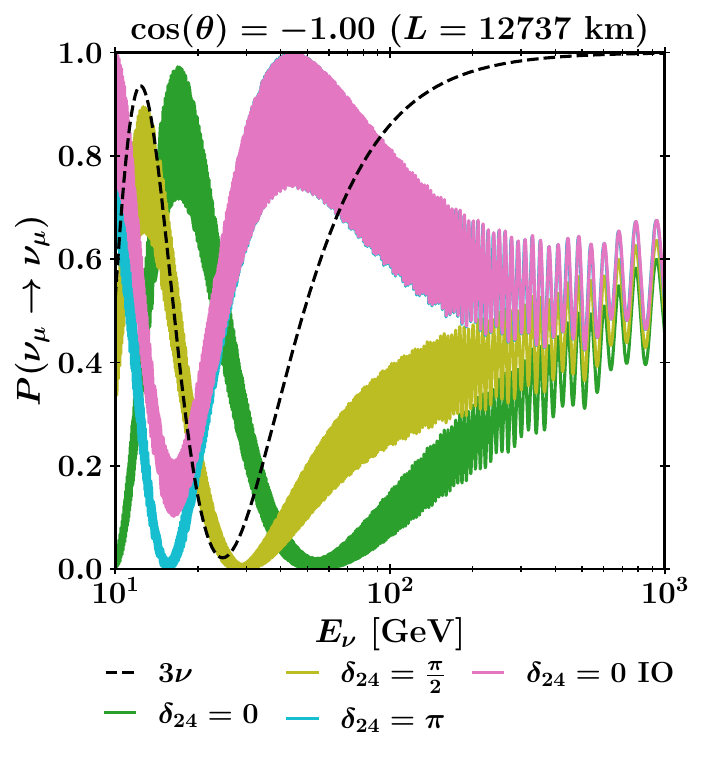}
  \caption{Muon neutrino survival probability as a function of $E_\nu$, for vertically up-going neutrinos ($\cos \theta = -1.0$). Normal ordering is assumed except when mentioned otherwise, with NuFit 5.2 values \cite{estebanFateHintsUpdated2020} for the standard oscillation parameters. The three-flavour model (dashed black curve) is compared with various 3$+$1 models all assuming $\Delta m^2_{41} = 1$ eV\tsup{2}. Left: different values of $\theta_{24}$ and $\theta_{34}$ ($s^2_{ij} = \sin^2\theta_{ij}$), with $\delta_{24}$ and other active-sterile mixing parameters fixed to 0. Right: different $\delta_{24}$ values, with $\sin^2 \theta_{24} = 0.07$, $\sin^2 \theta_{34} = 0.1$ and other active-sterile mixing parameters fixed to 0.}
  \label{f:sterile_1eV_1D_mixingangles}
\end{figure}

Instead, the analysis presented in this paper primarily focuses on the lower energy feature highlighted in figure \ref{f:sterile_1eV_2-dimensional_numu} (left). While the average effect of the presence of the sterile neutrino is an overall decrease of the $\nu_\mu$ survival probability compared to the three-flavour case, there is a notable exception. For the most vertical neutrinos, crossing both the mantle and the core ($\cos \theta < -0.838$), there is a region around $25$ GeV where less muon neutrinos disappear in the 3$+$1 case at the first $\nu_\mu \rightarrow \nu_\tau$ standard oscillation maximum. The effect of the $\theta_{24}$ and $\theta_{34}$ mixing angles (assuming $\delta_{24}=0$) on that feature is illustrated in the left panel of figure \ref{f:sterile_1eV_1D_mixingangles}. As previously observed, when $\theta_{24} \neq 0$, the amplitude of the survival minimum is reduced. $\theta_{34} \neq 0$ has the same effect. The degeneracy between $\theta_{24}$ and $\theta_{34}$ can be partially lifted by the lower (average) $\nu_\mu$ survival probability outside the minimum region seen in the case $\theta_{24} \neq 0$ but not for $\theta_{34} \neq 0$. In addition, this figure shows that when both active-sterile mixing angles are non-zero, the position of the minimum can be significantly shifted. This suggests that KM3NeT/ORCA could be particularly sensitive to simultaneously large values of $\theta_{24}$ and $\theta_{34}$. The right panel of figure \ref{f:sterile_1eV_1D_mixingangles} illustrates that the $\delta_{24}$ phase also affects the position of the minimum, with a limited effect on its amplitude (compared to the mixing angles). The known degeneracy between the NMO and the sign of $\cos \delta_{24}$ \cite{aielloSensitivityLightSterile2021} is also illustrated in figure \ref{f:sterile_1eV_1D_mixingangles} (right) as the curve obtained assuming IO and $\delta_{24} = 0$ overlaps the $\delta_{24} = \pi$ curve obtained in NO over most of the energy range considered. As all the effects mentioned here affect the $\nu_\mu \rightarrow \nu_\tau$ oscillations, degeneracies with the standard atmospheric parameters $\Delta m^2_{31}$ and $\theta_{23}$ are also expected. For instance, an apparent shift in $\Delta m^2_{31}$ could be accommodated by non-zero values for $\theta_{24}$ and $\theta_{34}$. 

\section{The KM3NeT/ORCA detector \label{s:det}}

The KM3NeT collaboration is building two water Cherenkov neutrino telescopes at the bottom of the Mediterranean Sea \cite{adrian-martinezLetterIntentKM3NeT2016}. The KM3NeT/ARCA (Astroparticle Research with Cosmics in the Abyss) detector, optimised for the detection of neutrinos from astrophysical sources in the TeV to PeV energy range, is located at a depth of about 3500 m, 80 km offshore Portopalo di Capo Passero in Sicily. The KM3NeT/ORCA (Oscillation Research with Cosmics in the Abyss) detector, optimised for the determination of the NMO through the measurement of oscillations of GeV to TeV atmospheric neutrinos crossing the Earth, is located 40 km offshore Toulon (France) at a depth of about 2500 m. Oscillation studies in the KM3NeT/ORCA energy range are well-suited to probe Beyond the Standard Model (BSM) hypotheses such as the 3$+$1 sterile neutrino model.\\

Both detectors collect the Cherenkov light induced by the relativistic charged particles emerging from neutrino interactions. They consist of 3-dimensional arrays of glass spheres named Digital Optical Modules (DOMs) \cite{aielloKM3NeTMultiPMTOptical2022a} housing 31 3-inch photomultiplier tubes (PMTs) each, arranged along flexible vertical structures called Detection Units (DUs) carrying 18 DOMs each. For KM3NeT/ORCA, the vertical spacing between DOMs is $9$ m. Each DU is anchored to the seabed and remains vertical due to the buoyancy of the DOMs and of a buoy installed at its top. The DUs are arranged following a cylindrical footprint, with a horizontal spacing of $20$ m. When completed, KM3NeT/ORCA will consist of 115 DUs, instrumenting a mass of sea water of around $7$ Mton.\\

When the pulse resulting from the detection of a Cherenkov photon by a PMT reaches a certain threshold, a \textit{hit} is generated. To save bandwidth, rather than digitizing the whole pulse, only the Time Over Threshold (ToT) of the pulse is recorded, along with the time of the hit and the PMT identifier. The hit information is sent to the onshore data processing center for online data filtering. More information about the data acquisition system of KM3NeT can be found in \cite{chiarusiKM3NeTDataAcquisition2023}. \textit{Events} are generated by trigger algorithms looking for causally-related hits on multiple DOMs. The trigger algorithms are designed to suppress background hits coming from dark counts (spontaneous electron emission from the photocathode of the PMTs) \cite{aielloCharacterisationHamamatsuPhotomultipliers2018}, bioluminescence \cite{haddockBioluminescenceSea2010, widderBioluminescencePelagicVisual2002a, aguzziInertialBioluminescenceRhythms2017}, and natural radioactivity of \tsup{40}K present in the sea water \cite{km3netcollaborationDependenceAtmosphericMuon2020}. The KM3NeT detectors are not sensitive to the detailed particle content of an event. Events are grouped into two topologies in KM3NeT/ORCA: \textit{track-like} and \textit{shower-like} events. Track-like events are caused by muons crossing the detector, encompassing $\nu_\mu$-CC interactions and $\nu_\tau$-CC interactions quickly followed by the decay of the $\tau$ lepton into a muon (branching ratio $\simeq 17.4\%$ \cite{particledatagroupReviewParticlePhysics2024}). Track-like events also include large amounts of down-going high-energy atmospheric muons which still reach the detectors despite the natural shielding provided by the sea water \cite{aielloAtmosphericMuonsMeasured2024}. GeV muons travel in a straight line and lose energy continuously at a minimum rate of about 0.2 GeV/m. Hence, the induced hits are observed on DOMs distributed evenly inside a cylinder, whose axis is the muon trajectory. The other neutrino interaction channels ($\nu_e$-CC, other $\nu_\tau$-CC, $\nu$-NC) create particle cascades (hadronic and/or electromagnetic), and are hence shower-like events. They appear as a localised burst of light in the detector, with hits distributed inside an ellipsoid whose longitudinal extent is 10 m at most \cite{adrian-martinezLetterIntentKM3NeT2016}. Based on the expected hit distributions, three trigger algorithms are applied in parallel: one looking for track-like events as sets of causally related hits within the volume of a cylinder, and two looking for shower-like events as sets of causally related hits within the volume of a sphere, with one of the two optimised for low-energy events.\\

Events are reconstructed assuming either a track or a shower model. More information about the track and shower reconstruction is given in refs. \cite{ofearraighFollowingLightNovel2024} and \cite{domiShowerReconstructionSterile2019a} respectively. Both procedures are iterative and use a maximum likelihood estimation based on the multi-dimensional probability density function of the photon arrival time on the PMTs.\\

To fulfill the physics goals of KM3NeT, good angular and energy resolutions must be achieved. Both require an accurate calibration of the detector. The KM3NeT detectors are huge infrastructures deployed in a changing environment which affects their performance over time. The position and orientation of the DOMs are measured every 10 minutes with an accuracy of 20 cm using an acoustic positioning system and sets of accelerometers and magnetometers located inside each DOM \cite{gatiusoliverDynamicalPositionOrientation2023}. The synchronisation of all PMTs with an onshore master clock is ensured through a custom implementation of the White Rabbit protocol \cite{serranoWhiteRabbitProject2009, aielloKM3NeTBroadcastOptical2023}. As delays can occur at multiple levels of the infrastructure, the time calibration must be achieved at various scales: between the PMTs of a given DOM (intra-DOM), between the DOMs of a given DU (inter-DOM), and between the DUs (inter-DU). The intra-DOM calibration uses coincident hits induced by \tsup{40}K decays \cite{melisInSituCalibrationKM3NeT2017}; the inter-DOM calibration uses dedicated LED flashers called nanobeacons \cite{aielloNanobeaconTimeCalibration2022}; and the inter-DU calibration relies on the maximization of the quality of the reconstructed tracks of atmospheric muons \cite{bailly-salinsAtmosphericMuonStudies2024, bailly-salinsTimePositionOrientation2023}.

\section{ORCA6 data sample and event selection \label{s:samplE_sel}}

Data were taken with the ORCA6 configuration from January 2020 to November 2021, with a total detector livetime of 633 days. After applying the data quality cuts presented in \cite{aielloMeasurementNeutrinoOscillation2024}, a total of 510 days of data taking remains, corresponding to a total exposure of 433 kton-years. This data sample was also used for other recently published ORCA6 oscillation analyses, comprising a standard neutrino oscillation analysis \cite{aielloMeasurementNeutrinoOscillation2024}, a study of tau neutrinos and unitarity \cite{aielloStudyTauNeutrinos2025}, searches for non-standard neutrino interactions \cite{aielloSearchNonstandardNeutrino2025a}, invisible neutrino decay \cite{aielloProbingInvisibleNeutrino2025a} and quantum decoherence \cite{aielloSearchQuantumDecoherence2025}.\\

In order to remove accidental triggers, pre-selection cuts on the number of hits and the reconstructed track likelihood of the events are applied \cite{aielloMeasurementNeutrinoOscillation2024}. In addition, only events reconstructed as up-going are kept. Even though only a small fraction of atmospheric muons are misreconstructed as up-going, they constitute close to $99.9\%$ of the accepted events at this stage, as the rate of reconstructed atmospheric muon events ($\approx 7$ Hz) exceeds that of neutrino events by 5 orders of magnitude.\\

The event selection and classification is performed using two dedicated sets of Boosted Decision Trees (BDTs), trained using Monte Carlo (MC) simulations (see section \ref{s:evt_distr}). Both BDTs are trained on (and applied to) events passing the pre-selection cuts only. The BDT models are built using variables related to the space and time distribution of the hits, as well as variables computed through the track and shower reconstruction chains, which are both applied to all triggered events. Each BDT model summarises the input features into a single output score. Details on the training and performances of the BDTs are given in \cite{aielloMeasurementNeutrinoOscillation2024}. The first BDT is used to separate neutrino events from the misreconstructed atmospheric muons events. It is able to identify poorly reconstructed atmospheric muons because they trigger more hits in the upper hemisphere of the DOMs and in the upper part of the DUs. Events with a high atmospheric muon BDT score are removed, resulting in a neutrino sample that contains 5828 events, with an expected muon contamination of a few percent. The goal of the second BDT is the separation between track-like and shower-like neutrino events. The resulting track sample is further divided in two classes: a \textit{high purity} track class with an expected atmospheric muon contamination around $0.1\%$, and around $95\%$ $\nu_\mu$-CC events; and a \textit{low purity} track class with a few percent muon contamination and around $90\%$ $\nu_\mu$-CC events. This selection leads to a total of 1868, 2002 and 1958 events in the high purity track, low purity track and shower classes respectively.\\

The track/shower separation gives some sensitivity to the flavour of the interacting neutrino, as events classified as tracks are mostly $\nu_\mu$-CC. However, around $45\%$ of the selected shower-like events are low-energy ($<10$ GeV) $\nu_\mu$-CC events, as short muon tracks cannot be easily distinguished from shower-like events. The low/high purity separation for track-like events increases the sensitivity to the standard oscillation parameters by isolating tracks with the best angular resolution in the high purity class \cite{aielloMeasurementNeutrinoOscillation2024}. For the eV-scale sterile neutrino search presented here, the same class definitions are kept, as the main feature giving sensitivity to the active-sterile mixing angles of interest is the first $\nu_\mu \rightarrow\nu_\tau$ oscillation maximum (see section \ref{s:oscprob}), the same used to measure $\theta_{23}$ and $\Delta m^2_{31}$ \cite{aielloMeasurementNeutrinoOscillation2024}. The reconstructed energy range for the two track classes is 2 GeV to 100 GeV. It is extended to 1 TeV for the shower class.\\

\section{Analysis method \label{s:methods}}

This analysis relies on comparing the observed numbers of reconstructed and selected events to the expected numbers, given a set of oscillation parameters of interest and nuisance parameters. This comparison is performed as a function of the reconstructed neutrino energy $E$ and the cosine of the zenith angle $\cos \theta$.

\subsection{Event distribution modeling and Monte Carlo simulations\label{s:evt_distr}}

To compute the expected number of events, the atmospheric neutrino flux is multiplied by the oscillation probabilities and weighted by the CC and NC interaction cross-sections, before accounting for the detector response. For the atmospheric neutrino flux, the year-averaged, solar minimum HKKM 2014 \cite{hondaAtmosphericNeutrinoFlux2015} flux tables computed for the Fréjus experiment site (located in Modane, 250 km away from the KM3NeT/ORCA site) are used. The OscProb software \cite{joao_coelho_2023_10104847} is used to compute the oscillation probabilities, and the Earth density is described using the Preliminary Reference Earth Model \cite{dziewonskiPreliminaryReferenceEarth1981}. The detector response is modeled through response matrices computed from MC simulations, briefly described in the following (see \cite{adrian-martinezLetterIntentKM3NeT2016} and \cite{aielloDeterminingNeutrinoMass2022} for more details). The generation of neutrino events is performed using the GENIE-based \cite{andreopoulosGENIENeutrinoMonte2010} software gSeaGen \cite{aielloGSeaGenKM3NeTGENIEbased2020}. Atmospheric muons are generated using the parametric generator MUPAGE \cite{carminatiAtmosphericMUonsPArametric2008}. The propagation of the Cherenkov photons induced along the path of the produced charged particles is performed using the Geant4-based \cite{agostinelliGeant4aSimulationToolkit2003} package KM3Sim \cite{tsirigotisHOUReconstructionSimulation2011}. For high-energy particles, a faster, custom KM3NeT package relying on probability density functions of the light arrival time is used instead of KM3Sim because of the large amounts of photons generated. The PMT response is simulated using another custom KM3NeT package which also includes the simulation of dark count and optical background, based on the trigger rates observed in the data for each PMT. From this point on, the output of the MC simulation chain is processed using the same trigger and event reconstruction algorithms (see section \ref{s:det}) and the same selection and classification procedure (see section \ref{s:samplE_sel}) as for data. A 4-dimensional response matrix is built for each event class and each neutrino interaction channel, mapping the true neutrino energy and cosine of the zenith angle to the reconstructed ones. The choice of the binning used to build the response matrices is described in \cite{aielloMeasurementNeutrinoOscillation2024}. The full chain to compute response matrices and expected event distributions is implemented through the custom KM3NeT software \textit{Swim} \cite{bourretNeutrinoOscillationsEarth2018, carreterocuencaNeutrinoOscillationsInvisible2024}.

\subsection{Nuisance parameters}

To model sources of systematic uncertainties, a total of 16 nuisance parameters are used in this analysis. Most are common to the standard oscillation analysis \cite{aielloMeasurementNeutrinoOscillation2024}:
\begin{itemize}
\item Uncertainties on the atmospheric neutrino flux shape and neutrino species ratios are modeled through corrections applied directly to the HKKM flux: on the ratio of up-going to horizontal neutrino ($\delta_\theta$), on the spectral index ($\delta_\sigma$), and on the ratios of, respectively, $\nu_\mu$ to $\bar{\nu}_\mu$ ($s_{\mu\bar{\mu}}$),  $\nu_e$ to $\bar{\nu}_e$ ($s_{e\bar{e}}$), and $\nu_e+\bar{\nu}_e$ to $\nu_\mu+\bar{\nu}_\mu$ ($s_{e\mu}$).
\item Uncertainties on neutrino cross-sections and on the selection and classification efficiencies are modeled through scaling factors affecting the normalisation of specific channels. The overall normalisation $f_\trm{all}$ scales all selected events. The normalisation of the NC ($\tau$-CC) events $f_\trm{NC}$ ($f_{\tau\trm{CC}}$) accounts for the uncertainty in modelling the NC interaction ($\tau$-CC) cross-section and event selection. The $f_\trm{HPT}$ and $f_\trm{S}$ normalisations are used for the relative normalisation of the high purity track and shower classes. A normalisation for the atmospheric muon background $f_{\mu}$ is also used. An additional scaling factor $f_\trm{HE}$ is used to scale high-energy events to account for the different assumptions made on light propagation by the two different light propagation software packages mentioned in section \ref{s:evt_distr}. This scaling is applied for NC events with true energy above 100 GeV and for CC events with true energy above 500 GeV.
\item Uncertainties on the water absorption and scattering lengths and the absolute PMT efficiency are accounted for through an energy scale parameter $E_s$. This parameter is applied as a shift in the true energy of the detector response. For more details and discussion about the implementation, see  \cite{chauStudyAtmosphericNeutrino2021}.
\end{itemize}

For the sterile neutrino search, $\Delta m^2_{31}$ and $\theta_{23}$ are also nuisance parameters. No prior uncertainty is assumed for $\theta_{23}$. $\Delta m^2_{31}$ is constrained using the values and uncertainties reported by the Daya Bay collaboration \cite{dayabaycollaborationPrecisionMeasurementReactor2023}. The Daya Bay measurement, obtained from $\nu_e$ disappearance, can be considered uncorrelated with the sterile neutrino search presented in this work, which relies mostly on the $\nu_\mu$ disappearance channel in a ``no $\nu_e$" approximation (see section \ref{s:res}). The other standard oscillation parameters, for which KM3NeT/ORCA has no sensitivity, are fixed to the NuFit 5.0 values with Super-Kamiokande data included in the global fit \cite{estebanFateHintsUpdated2020} given in table \ref{t:std_params}. Finally, the $\delta_{24}$ phase which, as shown in figure \ref{f:sterile_1eV_1D_mixingangles}, influences the position of the $\nu_\mu \rightarrow \nu_\tau$ standard oscillation maximum, is fitted without prior constraint. 

\begin{table}[H]
\centering
    \begin{tabular}{|l|c|c|}
      \hline
      \tbf{Parameter} & \tbf{NO} & \tbf{IO} \\
      \hline
      $\theta_{12}$ & $33.44^{\circ}$ & $33.44^{\circ}$ \\
      \hline
      $\theta_{13}$  & $8.57^{\circ}$ & $8.60^{\circ}$ \\     
      \hline
      $\Delta m_{21}^2$ [$10^{-5}~\mathrm{eV^2}$] & $7.42$ & 7.42 \\ 
      \hline
      $\delta_{\text{CP}}$ & $197^{\circ}$ & $282^{\circ}$ \\
      \hline
    \end{tabular}
    \caption{Values used for the non-fitted standard neutrino oscillation parameters, from NuFit 5.0 including Super-Kamiokande data \cite{estebanFateHintsUpdated2020}.}
    \label{t:std_params}
\end{table}

\subsection{Statistical methods}

In the following, given an observed number of events $n_{ij}$ of the class $i$ in bin $j$, the parameters of interest of the model $\vec{x}$ are determined through the minimisation of the following negative log-likelihood function:

\begin{equation}
\label{e:lik_BB}
\begin{aligned}
-2 \ln \mathcal{L}(\vec{x},\vec{\eta}) = 2& \sum_{i=1}^{N_\trm{classes}} \sum_{j=1}^{N_\trm{bins}} \left[ \beta_{ij}\mu_{ij}(\vec{x},\vec{\eta}) - n_{ij} + n_{ij} \ln\left(\frac{n_{ij}}{\beta_{ij}\mu_{ij}(\vec{x},\vec{\eta})}\right) + \frac{(\beta_{ij}-1)^2}{\sigma_{\beta_{ij}}^2}\right] \\+& \sum_{k=1} \left(\frac{\eta_k-\langle \eta_k\rangle}{\sigma_k}\right)^2\,,
\end{aligned}
\end{equation} 
where $\mu_{ij}(\vec{x},\vec{\eta})$ is the expected event distribution for the set of parameters of interest $\vec{x}$ and nuisance parameters $\vec{\eta}$. The $\beta_{ij}$ factors act as additional nuisance parameters and are introduced to account for the finite number of generated MC events used to build the response matrices, following the Barlow and Beeston light method \cite{barlowFittingUsingFinite1993, conwayIncorporatingNuisanceParameters2011} (see ref \cite{carreterocuencaNeutrinoOscillationsInvisible2024} for details). The term in square brackets corresponds to a Poissonian likelihood. The term in parentheses is a Gaussian penalisation term accounting for the prior uncertainties on the subset of constrained nuisance parameters: for a parameter $\eta_k$, $\langle \eta_k\rangle$ is the prior mean and $\sigma_k$ is its standard deviation.
In \textit{Swim}, the parameters of interest $\vec{x}$ and the nuisance parameters $\vec{\eta}$ are determined by minimising equation \ref{e:lik_BB} using the MIGRAD solver of the Minuit2 package \cite{ROOTMinuit2MnMigrad}.\\

The test statistic used to build the confidence interval of the parameters of interest around their best-fit values is the logarithm of the likelihood ratio between each point of the phase-space of interest $\vec{x}$ and the best-fit point $\vec{\wh{x}}$, written $-2\Delta \ln \mathcal{L}$ in the following. The confidence intervals of the parameters of interest are built in a frequentist way. Using Wilks' theorem \cite{wilksLargeSampleDistributionLikelihood1938}, and not a full Feldman-Cousins confidence interval derivation \cite{feldmanUnifiedApproachClassical1998} due to computing constraints, the allowed region of the parameter space given a certain confidence level is obtained by comparing the value of $-2\Delta \ln \mathcal{L}$ with the corresponding quantile of a $\chi^2$ distribution with two degrees of freedom, as two is the dimension of the parameter space of interest.

\subsection{Fitting procedure}

For each point of the phase space of interest, equation \ref{e:lik_BB} is minimised using two starting points for four nuisance parameters, to avoid getting stuck in local minima: normal and inverted ordering ($\Delta m^2_{31} = \{ 2.541\times10^{-3}, -2.496\times10^{-3}\}~\mathrm{eV^2}$); $\theta_{23}$ lower and upper octant ($\theta_{23} =\{40^{\circ}, 50^{\circ}\}$); energy scale $E_s$ below and above 1.0 ($E_s=\{0.95, 1.05\}$); and $\delta_{24}$ below and above $180^{\circ}$  ($\delta_{24} =\{90^{\circ}, 270^{\circ}\}$). This produces a total of 16 distinct starting points. The two start values for $\Delta m_{31}^2$, $\theta_{23}$ and $E_s$ are common to the standard oscillation analysis of ORCA6 \cite{aielloMeasurementNeutrinoOscillation2024}. For $\delta_{24}$, the use of two start values is motivated by the observation of local minima when computing the $-2\Delta \ln \mathcal{L}$ profile of that parameter. In each fit, the parameter space is restricted to the NMO, $\theta_{23}$ octant and $(E_s-1)$ sign corresponding to the start value. No limit is put on $\delta_{24}$.

\section{Results \label{s:res}}

The ORCA6 data sample is analysed under the $\Delta m_{41}^2 = 1$ eV\tsup{2} hypothesis; the results hold for $\Delta m_{41}^2 > 1$ eV\tsup{2} for which faster oscillations are unobservable. Simultaneous constraints on the mixing of the $\nu_\mu$ and $\nu_\tau$ states with the sterile state are obtained by scanning the $(|U_{\mu 4}|^2, |U_{\tau 4}|^2)$ space, with $U_{e4}=0$ (i.e. $\theta_{14}=\delta_{14}=0$). In that case, the magnitude of the $U_{\mu 4}$ and $U_{\tau 4}$ mixing elements is expressed as:
\begin{align}
	|U_{\mu 4}|^2 &= \, \sin^2 \theta_{24}\\
	|U_{\tau 4}|^2 &= \, \cos^2 \theta_{24} \sin^2 \theta_{34}\,.
\end{align}

For each fit $\delta_{24}$ is set free. To obtain the allowed region for the parameters of interest at a given confidence level, a 2-dimensional scan is performed by sampling the $(|U_{\mu 4}|^2, |U_{\tau 4}|^2)$ phase space uniformly in log scale, using 27 points between $1\times10^{-3}$ and $0.5$ (excluded) for both $|U_{\mu 4}|^2$ and $|U_{\tau 4}|^2$, with an additional point at 0. 

\subsection{Best-fit point}

The position of the best-fit point $\vec{\wh{x}}$ is computed independently of the grid scan: fits with free $U_{\mu 4}$ and $U_{\tau 4}$ (thus free $\theta_{24}$ and $\theta_{34}$) are performed. To avoid falling in potential local minima in the $(|U_{\mu 4}|^2, |U_{\tau 4}|^2)$ space, three pairs of starting values are used (in addition to the $16$ sets of starting values already introduced): $(0.0, 0.0)$, $(0.5, 0.0)$ and $(0.0, 0.5)$. This makes a total of 48 fits to determine the best-fit point, found at:

\begin{equation}
	\begin{aligned}
	|U_{\mu 4}|^2 &=\, 6.89 \times 10^{-2}\\
	|U_{\tau 4}|^2 &=\, 2.35 \times 10^{-4}\,,
	\end{aligned}
	\label{e:BF}
\end{equation}
which in terms of mixing angles corresponds to $\theta_{24} = 15.2 \degree$ and $\theta_{34} = 0.9 \degree$. The values of the nuisance parameters obtained at the best-fit point are compared with their assumed prior uncertainty in table \ref{t:BF-values}. The most significant deviations with respect to the results of the standard oscillations best fit, reported in table 4 of \cite{aielloMeasurementNeutrinoOscillation2024}, are for the overall normalisation $f_{\trm{all}}$, with $1.27^{+0.15}_{-0.14}$ in the sterile fit versus $1.11^{+0.14}_{-0.13}$ in the three-flavour fit, and the energy scale $E_s$ with $1.10^{+0.07}_{-0.07}$ versus $1.03^{+0.11}_{-0.08}$. The higher overall normalisation in the sterile case is expected, as the average effect of the sterile neutrino is to decrease the number of neutrinos of other flavours. The difference in energy scale can be explained by the prior on $\Delta m_{31}^2$ used for the current sterile analysis but not for the standard oscillation analysis. Indeed, this prior strongly restricts the fit to values of $|\Delta m_{31}|^2$ close to $2.5\times 10^{-3}$  eV\tsup{2} in the sterile analysis, while a value of $2.18\times 10^{-3}$  eV\tsup{2} is found to match the data better in the standard oscillation analysis, where no external constraint on that parameter is imposed. Thus, in the sterile fit, the larger $|\Delta m_{31}|^2$ value imposed means that the $L/E$ ratio should be scaled down to equivalently describe the event distributions. This means the fit favours a higher scaling of the energy, explaining the higher $E_s$ seen in the sterile fit. The needed shift in energy induced by the larger $|\Delta m_{31}|^2$ value imposed in this analysis is also partially accommodated by the values of the sterile parameters (see section \ref{s:oscprob}). For vertical up-going neutrinos, the position of the first $\nu_\mu \rightarrow \nu_\tau$ maximum is shifted by 2.4$\%$ (from 24.7 to 25.3 GeV) when comparing the oscillation probabilities of the 3$+$1 model at the best-fit point with the ones of the standard three-flavour model.

\begin{table}[H]
    \centering
\begin{tabular}{|c|c|c|c|}
 \hline
   Parameter & Nominal value $\pm$ uncertainty& Best Fit & Post-fit uncertainty   \\ \hline
   $\delta_\theta$ & 0.00 $\pm$ 0.02 & $-$0.00& $-$0.02/+0.02 \\ \hline 
   $\delta_{\gamma}$ & 0.00 $\pm$ 0.30 & $-$0.00& $-$0.03/+0.03 \\ \hline  
   $s_{e \bar{e}}$ & 0.00 $\pm$ 0.07 & 0.01& $-$0.07/+0.07 \\ \hline
   $s_{\mu \bar{\mu}}$ & 0.00 $\pm$ 0.05 & 0.00& $-$0.05/+0.05 \\ \hline
   $s_{e\mu}$ & 0.00 $\pm$ 0.02 & $-$0.00& $-$0.02/+0.02 \\ \hline
   $f_{\tau\trm{CC}}$ & 1.00 $\pm$ 0.20& 0.90& $-$0.18/+0.18 \\ \hline
   $f_{\trm{NC}}$ & 1.00 $\pm$ 0.20 & 0.85& $-$0.19/+0.19 \\ \hline
   $E_s$ & 1.00 $\pm$ 0.09 & 1.10& $-$0.07/+0.07 \\ \hline
   $f_{\trm{HE}}$ & $1.00 \pm 0.50$ & 1.57& $-$0.29/+0.32 \\ \hline   
   $f_{\trm{HPT}}$ & 1.00& 0.92& $-$0.04/+0.04 \\ \hline
   $f_{\trm{S}}$ & 1.00& 0.89& $-$0.06/+0.06 \\ \hline
   $f_{\mu}$ & 1.00& 0.40& $-$0.31/+0.36 \\ \hline
   $f_{\trm{all}}$ & 1.00& 1.27& $-$0.14/+0.15 \\ \hline
   $\Delta m_{31}^2$ [$10^{-3}$ eV\tsup{2}] & $2.541 \pm 0.060$  & 2.521 & $-$0.059/+0.059 \\ \hline
   $\theta_{23}$ [\textdegree] & 49.2 & 44.1 & $-$4.1/+6.4 \\ \hline  
   $\delta_{24}$ [\textdegree] & 0 & 322 & $-$322/+38 \\ 
\hline
\end{tabular}
    \caption{Best-fit values and post-fit uncertainties at $68\%$ CL of the nuisance parameters from the fit of ORCA6 data to $U_{\mu 4}$ and $U_{\tau 4}$ with $\Delta m_{41}^2 = 1$ eV\tsup{2}. The fit is done in both mass orderings, and the best fit is found in NO.}
       \label{t:BF-values}
\end{table}

\subsection{\boldmath $(|U_{\mu 4}|^2, |U_{\tau 4}|^2)$ scan results}

\begin{figure}[h!]
  \centering
  \includegraphics[width=0.554\linewidth, trim={0.5in 0 0.3in 0}, clip=True]{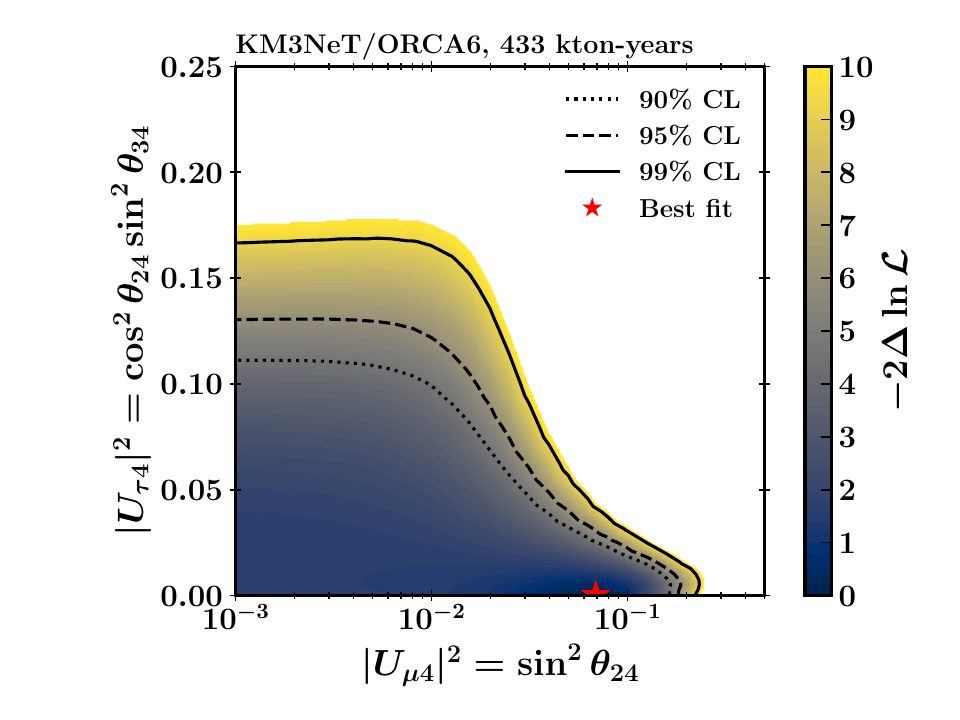}
  \includegraphics[width=0.436\linewidth, trim={0.7in 0 1.3in 0}, clip=True]{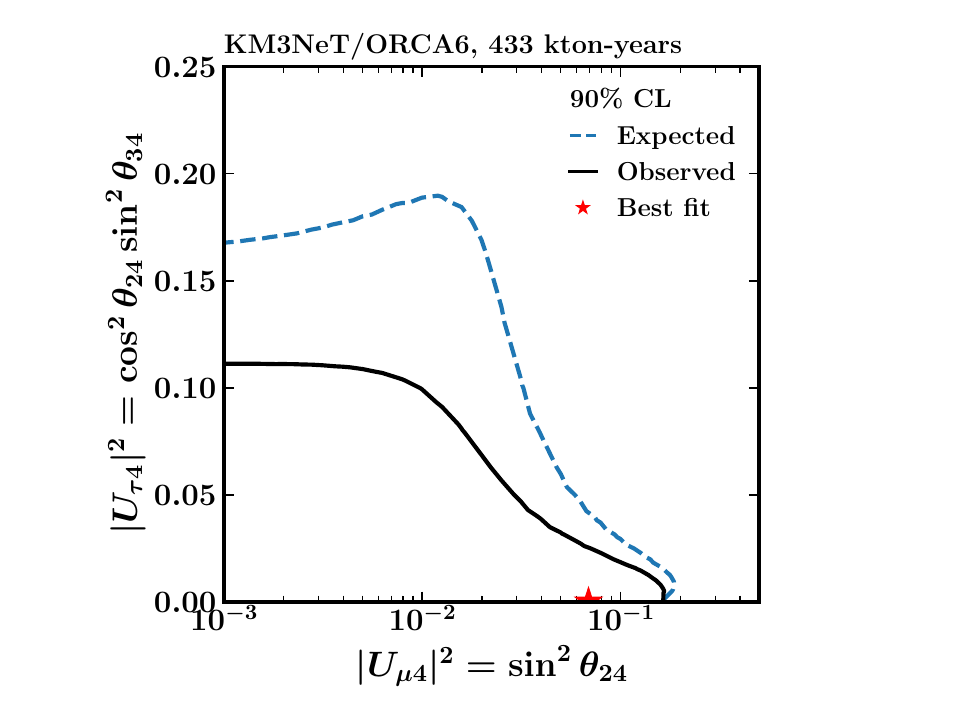}
  \vspace*{-8mm}
  \caption{\textbf{Left}: Log-likelihood ratio map obtained from the ORCA6 data sample over the $|U_{\mu 4}|^2$ and $|U_{\tau 4}|^2$ phase space. $-2\Delta \log \mathcal{L}$ is not displayed when higher than $10$ units. The black lines show upper limits of the allowed parameter space at various confidence levels. \textbf{Right}: Comparison of the observed and expected upper limits at $90\%$ CL on the $U_{\mu 4}$ and $U_{\tau 4}$ mixing elements for ORCA6. The excluded regions are on the top right side of the contours.}
  \label{f:unblind-LLR}
\end{figure}

The log-likelihood ratio map resulting from the 2-dimensional grid scan, as well as upper limits of the allowed parameter space at confidence levels (CL) of $90\%$, $95\%$ and $99\%$, are shown in figure \ref{f:unblind-LLR} (left). A linear scale is used for the $y$ axis so that the best fit, which is outside the scanned range, is visible. The $(0,0)$ point, with a value $-2\Delta \ln \mathcal{L}(0,0) = 1.76$, is within the allowed region for the three CLs considered. This means that the ORCA6 data is fully compatible with the standard model. When profiling over the other mixing element, the upper limits at a $90\%$ CL are found to be:
\begin{equation}
	\begin{aligned}
		|U_{\mu 4}|^2 &< \, 0.138\\
		|U_{\tau 4}|^2 &< \, 0.076\,.
	\end{aligned}
\end{equation}	

\begin{figure}[h!]
  \centering
  \includegraphics[width=1.0\linewidth, trim={0 0.5cm 0 0}, clip=True]{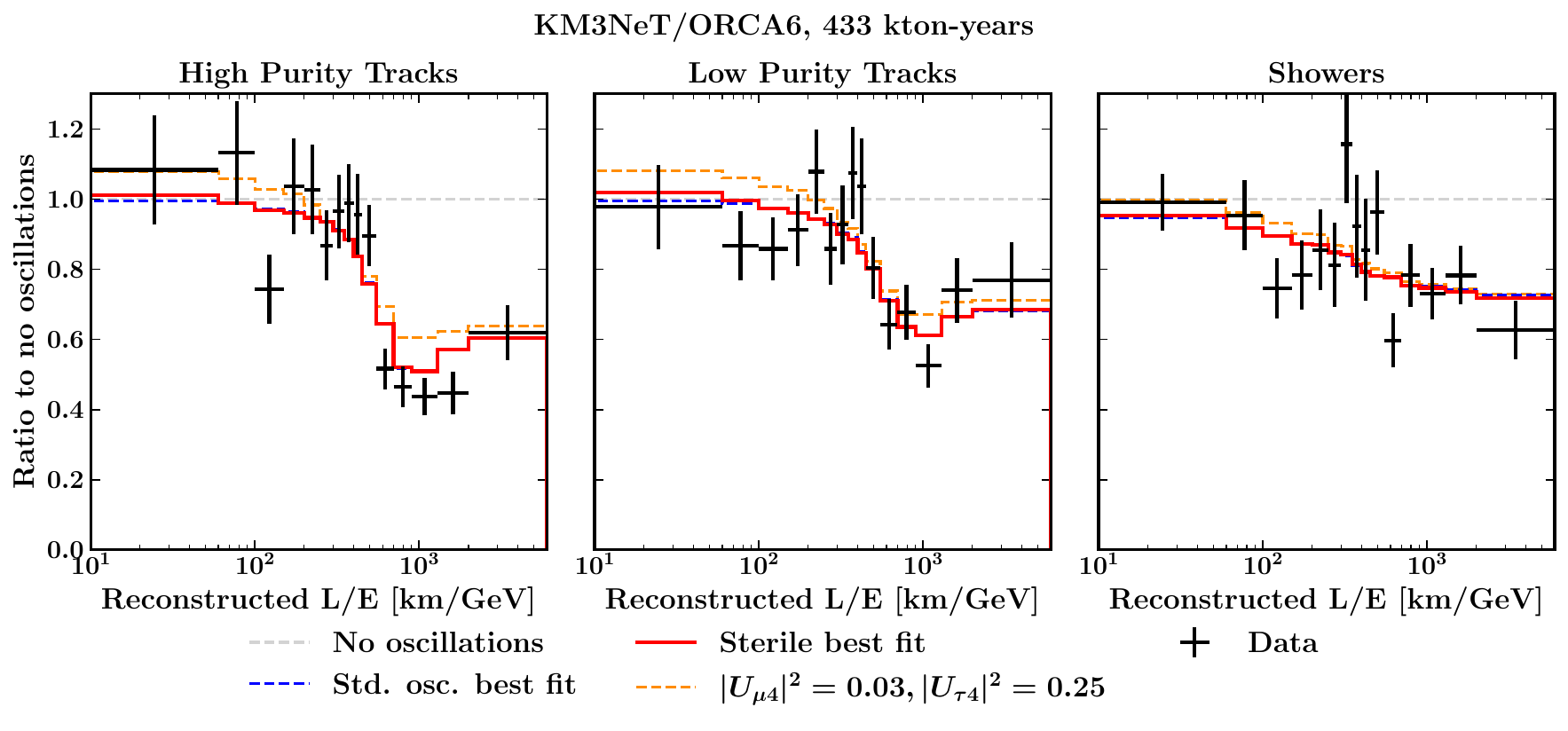}
  \vspace*{-5mm}  
  \caption{Event distributions in $L/E$ for each class of the ORCA6 data (black), compared with the model prediction at the best-fit point in the eV sterile neutrino case (red) and in the standard oscillation case (dashed blue). The dashed orange curve shows the model prediction for an arbitrary point in the $(|U_{\mu 4}|^2, |U_{\tau 4}|^2)$ phase space.}
  \label{f:unblind-BF-LoE}
\end{figure}

\begin{figure}[h!]
  \centering
  \includegraphics[width=0.8\linewidth, trim={0 0.35cm 0 0}, clip=True]{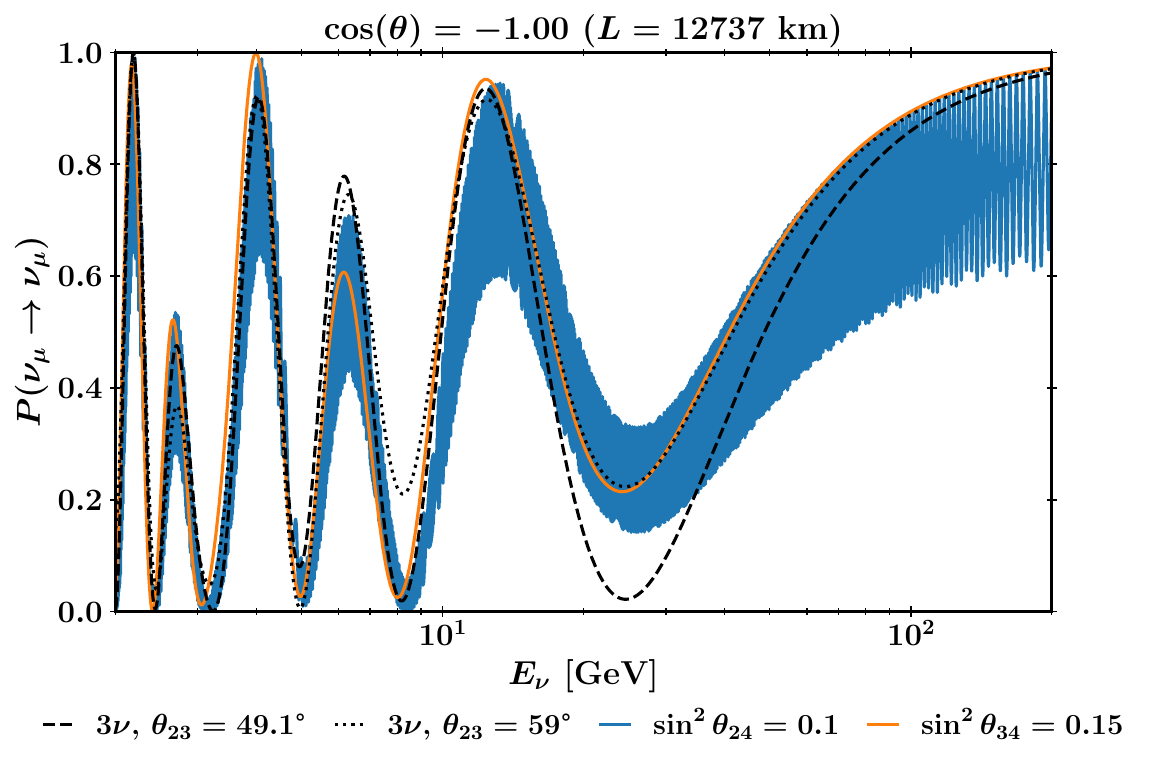}  
  \caption{Muon neutrino survival probability as a function of the neutrino energy under various scenarios for $\cos \theta = -1.0$. The three-flavour standard oscillation model assuming the nominal value for $\theta_{23}=49.1\degree$ (dashed black) is compared with the three-flavour model with $\theta_{23}=59\degree$ (dotted black) and with the 3$+$1 models with $\sin^2 \theta_{24} = 0.1$ (blue) and $\sin^2 \theta_{34} = 0.15$ (orange), for $\Delta m_{41}^2=1$ eV\tsup{2}. All other sterile parameters are set to 0.}
  \label{f:numu_disap_t23}
\end{figure}

In figure \ref{f:unblind-LLR} (right), the $90\%$ CL upper limits in $|U_{\mu 4}|^2$ and $|U_{\tau 4}|^2$ obtained by fitting the ORCA6 433 kton-years data set are compared with the ones obtained by fitting an Asimov MC data set generated using the data best-fit point values for the parameters of interest and the nuisance parameters (reported in table \ref{t:BF-values}). This comparison shows that a larger region of the phase space is excluded when fitting the data, compared to the one expected from the Asimov sensitivity study. The stricter upper limits obtained when fitting the data can be understood by looking at the one-dimensional reconstructed $L/E$ event distributions of figure \ref{f:unblind-BF-LoE}, shown for the three event classes. Both the data events and  model best-fit point for this analysis are shown. The standard oscillation best fit is also shown for reference. It is barely distinguishable from the sterile best fit. The standard oscillation pattern is clearly visible in the high purity (and to a lower extent, low purity) track class. However, it is hardly distinguishable in the shower class, due to the mixture of flavours in this class. On the high purity tracks distribution, the data underfluctuates with respect to the model at the oscillation dip (corresponding to the first $\nu_\mu \rightarrow \nu_\tau$ maximum)  at $L/E \sim \mathcal{O}(10^3)$ km/GeV. This is a crucial feature of the event distribution studied here as it constrains the model to give the deepest dip possible, which strongly favours maximal mixing for $\theta_{23}$ (as already reported in \cite{aielloMeasurementNeutrinoOscillation2024}). It also puts strong constraints on the active-sterile mixing angles $\theta_{24}$ and $\theta_{34}$, with particularly strict limits on $\theta_{34}$. Indeed, the effect of $\theta_{34}$ on the $P(\nu_\mu \rightarrow \nu_\mu)$ first maximum is very similar to the effect of a $\theta_{23}$ value far from maximal mixing, as illustrated in figure \ref{f:numu_disap_t23}, while $\theta_{24}$ lowers the survival probability outside the oscillation dip region. This explains why high values of $\theta_{34}$ are strongly rejected. For illustration, an arbitrarily high value of $|U_{\tau 4}|^2$ is chosen to draw the dashed orange curve in figure \ref{f:unblind-BF-LoE}, showing that higher $\theta_{34}$ makes the oscillation dip shallower, and therefore even further away from the data. 

\subsection{Comparison with existing measurements}

\begin{figure}[h!]
  \centering
  \includegraphics[width=0.7\linewidth, trim={1.2cm 0 0.8cm 0}, clip=True]{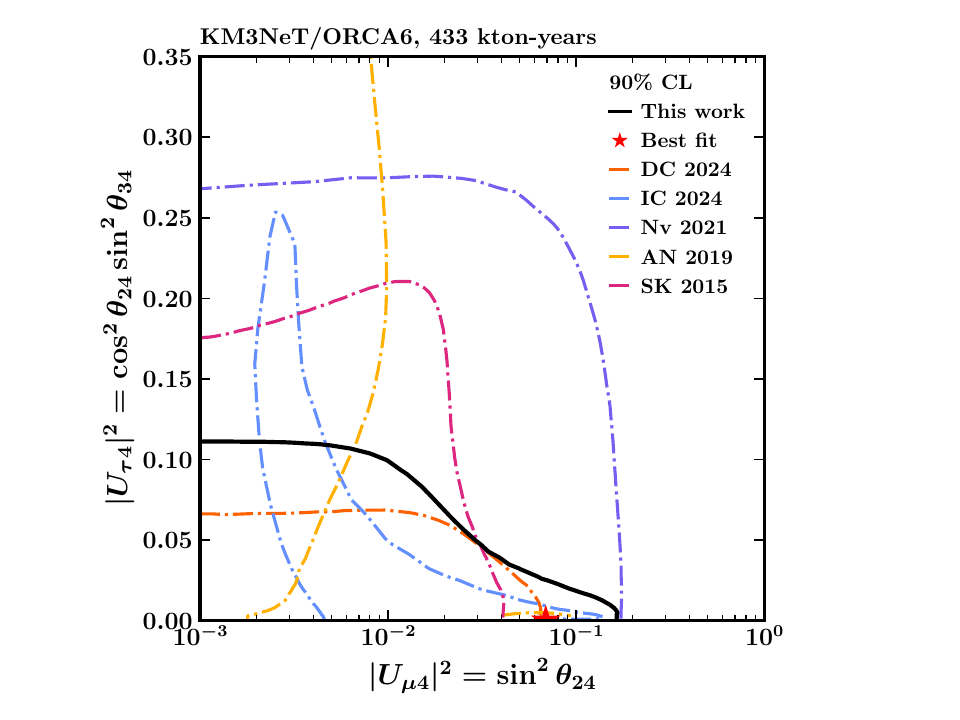}
  \caption{Comparison of the upper limits at $90\%$ CL on $|U_{\mu 4}|^2$ and $|U_{\tau 4}|^2$ obtained by ORCA6 with previous measurements from  Super-Kamiokande (SK) \cite{super-kamiokandecollaborationLimitsSterileNeutrino2015}, NOvA (Nv) \cite{thenovacollaborationSearchActiveSterileAntineutrino2021}, ANTARES (AN) \cite{albertMeasuringAtmosphericNeutrino2019}, IceCube (IC) \cite{abbasiExplorationMassSplitting2024a} and DeepCore (DC) \cite{icecubecollaborationSearchLightSterile2024}. The excluded regions are on the top right side of the contours.}
  \label{f:unblind-ext}
\end{figure}

The ORCA6 upper limits at $90\%$ CL are compared with the results of other experiments in figure \ref{f:unblind-ext}. Table \ref{t:ext_config} summarizes how each analysis treats the 3+1 and standard oscillation parameters relevant for this measurement. All consider eV-scale sterile neutrinos, but do not report exactly the same $\Delta m_{41}^2$ validity range. The IceCube result is obtained by profiling over $\Delta m_{41}^2$ between 0.1 eV\textsuperscript{2} and 50 eV\textsuperscript{2}, with a best fit found at 5.0 eV\textsuperscript{2}. All analyses considered here treat $\delta_{24}$ as a free nuisance parameter, with the exception of Super-Kamiokande and IceCube. The Super-Kamiokande analysis fixes $\delta_{24}=0$, leading to more restrictive limits than with a free $\delta_{24}$. The IceCube analysis relies on the TeV disappearance feature which is less sensitive to $\delta_{24}$. It uses $\delta_{24}=\pi$, as this choice was found to yield the most conservative limits. Regarding the mass ordering, only the ORCA6 and the ANTARES analyses consider both orderings. Finally, due to the partial degeneracies between $\theta_{23}$ and the $\theta_{24}$ and $\theta_{34}$ mixing angles (see figure \ref{f:numu_disap_t23}), analyses lead with no constraint on this parameter (as done in this work, and for the DeepCore and ANTARES analysis) yield more conservative limits than the other analyses.

\begin{table}[h!]
    \centering
\begin{tabular}{|c|c|c|c|c|c|}
 \hline
   Experiment & $\Delta m_{41}^2$ range (eV\tsup{2}) & $\delta_{24}$ & Mass ordering & $\Delta m_{31}^2$ & $\theta_{23}$ \\ \hline
   This work & $\geq 1.0$ & Free & Both & Prior from \cite{dayabaycollaborationPrecisionMeasurementReactor2023} & Free \\ \hline 
   DeepCore \cite{icecubecollaborationSearchLightSterile2024} & $\geq 1.0$ & Free & NO only & \multicolumn{2}{|c|}{Free} \\ \hline  
   IceCube \cite{abbasiExplorationMassSplitting2024a} & Profiled in $[0.1, 50]$ & $=\pi$ & Unknown & \multicolumn{2}{|c|}{Unknown} \\ \hline
   NOvA \cite{thenovacollaborationSearchActiveSterileAntineutrino2021} & $[0.05, 0.5]$ & Free & NO only & \multicolumn{2}{|c|}{Fixed from \cite{particledatagroupReviewParticlePhysics2018}} \\ \hline
   ANTARES \cite{albertMeasuringAtmosphericNeutrino2019} & $\geq 0.5$ & Free & Both & \multicolumn{2}{|c|}{Free} \\ \hline
   SK \cite{super-kamiokandecollaborationLimitsSterileNeutrino2015} & $\geq 0.1$ & $=0$ & NO only & \multicolumn{2}{|c|}{Prior from \cite{t2kcollaborationPreciseMeasurementNeutrino2014}} \\ 
\hline
\end{tabular}
    \caption{Summary of the configurations used for the analyses compared in figure \ref{f:unblind-ext}. The assumed $\Delta m_{41}^2$ validity range, and the treatment of $\delta_{24}$, the mass ordering, $\Delta m_{31}^2$ and $\theta_{23}$ in the fits are reported. Other oscillation parameters do not have a significant impact on the $|U_{\mu 4}|^2$ and $|U_{\tau 4}|^2$ measurement.}
       \label{t:ext_config}
\end{table}

The ORCA6 limits reported in figure \ref{f:unblind-ext} are competitive with the existing results, especially concerning the limit on the $\nu_\tau$ mixing with the sterile state: only DeepCore, using $7.5$ years of data, puts a stronger constraint on $|U_{\tau 4}|^2$. For $|U_{\mu 4}|^2$, ANTARES, IceCube, DeepCore and Super-Kamiokande all put stronger constraints than ORCA6, but these results were obtained with several years of data-taking from complete detectors, while the ORCA6 measurement was obtained with 1.4 years of data recorded with a detector that was $5\%$ of its final size. The performance of KM3NeT/ORCA is due to its good sensitivity  to $\nu_\mu$ in the $20-30$ GeV region which allows for the monitoring of the first $\nu_\mu \rightarrow \nu_\tau$ standard oscillation maximum. Other detectors have to rely on weaker effects at lower energy (e.g. Super-Kamiokande) or are optimised for slightly higher energies (e.g. ANTARES).

\section{Conclusion \label{s:ccl}}

The first search for an eV-scale sterile neutrino performed with the KM3NeT/ORCA detector is presented in this article. In this analysis, a data sample collected with 6 detection units ($5\%$ of the complete detector size) is used, corresponding to an exposure of 433 kton-years and containing 5828 neutrino candidates. Constraints on the mixing of a hypothetical eV-scale sterile neutrino with the $\nu_\mu$ and $\nu_\tau$ states are established for $\Delta m^2_{41} \geq 1$ eV$^2$. The results are compatible with the three-flavour standard model, and the upper limits obtained on $|U_{\mu 4}|^2$ and $|U_{\tau 4}|^2$ are compatible with the results reported by other experiments.\\ 

This result, although based on a limited detector size and statistics, highlights the significant potential of KM3NeT/ORCA. More specifically for eV-scale sterile neutrino searches, it should be possible to probe the sterile-induced TeV resonance on $\bar{\nu}_\mu$ disappearance with the final configuration of KM3NeT/ORCA.

\acknowledgments

The authors acknowledge the financial support of:
KM3NeT-INFRADEV2 project, funded by the European Union Horizon Europe Research and Innovation Programme under grant agreement No 101079679;
Funds for Scientific Research (FRS-FNRS), Francqui foundation, BAEF foundation.
Czech Science Foundation (GA\v{C}R 24-12702S);
Agence Nationale de la Recherche (contract ANR-15-CE31-0020), Centre National de la Recherche Scientifique (CNRS), Commission Europ\'eenne (FEDER fund and Marie Curie Program), LabEx UnivEarthS (ANR-10-LABX-0023 and ANR-18-IDEX-0001), Paris \^Ile-de-France Region, Normandy Region (Alpha, Blue-waves and Neptune), France,
The Provence-Alpes-C\^ote d'Azur Delegation for Research and Innovation (DRARI), the Provence-Alpes-C\^ote d'Azur region, the Bouches-du-Rh\^one Departmental Council, the Metropolis of Aix-Marseille Provence and the City of Marseille through the CPER 2021-2027 NEUMED project,
The CNRS Institut National de Physique Nucl\'eaire et de Physique des Particules (IN2P3);
Shota Rustaveli National Science Foundation of Georgia (SRNSFG, FR-22-13708), Georgia;
This research was funded by the European Union (ERC MuSES project No 101142396); 
The General Secretariat of Research and Innovation (GSRI), Greece;
Istituto Nazionale di Fisica Nucleare (INFN) and Ministero dell'Universit{\`a} e della Ricerca (MUR), through PRIN 2022 program (Grant PANTHEON 2022E2J4RK, Next Generation EU) and PON R\&I program (Avviso n. 424 del 28 febbraio 2018, Progetto PACK-PIR01 00021), Italy; IDMAR project Po-Fesr Sicilian Region az. 1.5.1; A. De Benedittis, W. Idrissi Ibnsalih, M. Bendahman, A. Nayerhoda, G. Papalashvili, I. C. Rea, A. Simonelli have been supported by the Italian Ministero dell'Universit{\`a} e della Ricerca (MUR), Progetto CIR01 00021 (Avviso n. 2595 del 24 dicembre 2019); KM3NeT4RR MUR Project National Recovery and Resilience Plan (NRRP), Mission 4 Component 2 Investment 3.1, Funded by the European Union -- NextGenerationEU,CUP I57G21000040001, Concession Decree MUR No. n. Prot. 123 del 21/06/2022;
Ministry of Higher Education, Scientific Research and Innovation, Morocco, and the Arab Fund for Economic and Social Development, Kuwait;
Nederlandse organisatie voor Wetenschappelijk Onderzoek (NWO), the Netherlands;
The grant "AstroCeNT: Particle Astrophysics Science and Technology Centre", carried out within the International Research Agendas programme of the Foundation for Polish Science financed by the European Union under the European Regional Development Fund; The program: 'Excellence initiative-research university' for the AGH University in Krakow; The ARTIQ project: UMO-2021/01/2/ST6/00004 and ARTIQ/0004/2021;
Ministry of Education and Scientific Research, Romania;
Slovak Research and Development Agency under Contract No. APVV-22-0413; Ministry of Education, Research, Development and Youth of the Slovak Republic;
MCIN for PID2021-124591NB-C41, -C42, -C43 and PDC2023-145913-I00 funded by MCIN/AEI/10.13039/501100011033 and by "ERDF A way of making Europe", for ASFAE/2022/014 and ASFAE/2022 /023 with funding from the EU NextGenerationEU (PRTR-C17.I01) and Generalitat Valenciana, for Grant AST22\_6.2 with funding from Consejer\'{\i}a de Universidad, Investigaci\'on e Innovaci\'on and Gobierno de Espa\~na and European Union - NextGenerationEU, for CSIC-INFRA23013 and for CNS2023-144099, Generalitat Valenciana for CIDEGENT/2020/049, CIDEGENT/2021/23, CIDEIG/2023/20, ESGENT2024/24, CIPROM/2023/51, GRISOLIAP/2021/192 and INNVA1/2024/110 (IVACE+i), Spain;
Khalifa University internal grants (ESIG-2023-008, RIG-2023-070 and RIG-2024-047), United Arab Emirates;
The European Union's Horizon 2020 Research and Innovation Programme (ChETEC-INFRA - Project no. 101008324).

Views and opinions expressed are those of the author(s) only and do not necessarily reflect those of the European Union or the European Research Council. Neither the European Union nor the granting authority can be held responsible for them.

\bibliographystyle{JHEP}
\bibliography{KM3NeT_sterile_paper.bib}

@article{aartsenSearchingEVscaleSterile2020,
  title = {Searching for {{eV-scale}} Sterile Neutrinos with Eight Years of Atmospheric Neutrinos at the {{IceCube}} Neutrino Telescope},
  collaboration = "IceCube",
  year = {2020},
  journal = {Phys. Rev. D},
  volume = {102},
  number = {5},
  primaryclass = {hep-ex, physics:hep-ph},
  pages = {052009},
  issn = {2470-0010, 2470-0029},
  doi = {10.1103/PhysRevD.102.052009},
  urldate = {2023-12-01},
  archiveprefix = {arXiv},
}

@techreport{aceroWhitePaperLight2023,
  title = {White {{Paper}} on {{Light Sterile Neutrino Searches}} and {{Related Phenomenology}}},
  author = {Acero, M. A. and Arg{\"u}elles, C. A. and Hostert, M. and Kalra, D. and Karagiorgi, G. and Kelly, K. J. and Littlejohn, B. and Machado, P. and Pettus, W. and Toups, M. and {Ross-Lonergan}, M. and Sousa, A. and Surukuchi, P. T. and Wong, Y. Y. Y. and Abdallah, W. and Abdullahi, A. M. and Akutsu, R. and {Alvarez-Ruso}, L. and Alves, D. S. M. and Aurisano, A. and Balantekin, A. B. and Berryman, J. M. and {Bert{\'o}lez-Mart{\'i}nez}, T. and Brunner, J. and Blennow, M. and Bolognesi, S. and Borusinski, M. and Cianci, D. and Collin, G. and Conrad, J. M. and Crow, B. and Denton, P. B. and Duvall, M. and {Fern{\'a}ndez-Martinez}, E. and Fong, C. S. and Foppiani, N. and Forero, D. V. and Friend, M. and {Garc{\'i}a-Soto}, A. and Giganti, C. and Giunti, C. and Gandhi, R. and Ghosh, M. and Hardin, J. and Heeger, K. M. and Ishitsuka, M. and Izmaylov, A. and Jones, B. J. P. and Jordan, J. R. and Kamp, N. W. and Katori, T. and Kim, S. B. and Koerner, L. W. and Lamoureux, M. and Lasserre, T. and Leach, K. G. and Learned, J. and Li, Y. F. and Link, J. M. and Louis, W. C. and Mahn, K. and Meyers, P. D. and Maricic, J. and Marko, D. and Maruyama, T. and Mertens, S. and Minakata, H. and Mocioiu, I. and Mooney, M. and Moulai, M. H. and Nunokawa, H. and {Ochoa-Ricoux}, J. P. and Oh, Y. M. and Ohlsson, T. and P{\"a}s, H. and Pershey, D. and Robertson, R. G. H. and {Rosauro-Alcaraz}, S. and Rott, C. and Roy, S. and Salvado, J. and Scott, M. and Seo, S. H. and Shaevitz, M. H. and Smiley, M. and Spitz, J. and Stachurska, J. and Thakore, T. and Ternes, C. A. and Thompson, A. and Tseng, S. and Vogelaar, B. and Weiss, T. and Wendell, R. A. and Wright, T. and Xin, Z. and Yang, B. S. and Yoo, J. and Zennamo, J. and Zettlemoyer, J. and Zornoza, J. D. and Ahmad, S. and {Basto-Gonzalez}, V. S. and Bowden, N. S. and Ca{\~n}as, B. C. and Caratelli, D. and Chang, C. V. and Chen, C. and Classen, T. and Convery, M. and Davies, G. S. and Dennis, S. R. and Djurcic, Z. and Dorrill, R. and Du, Y. and Evans, J. J. and Fahrendholz, U. and Formaggio, J. A. and Foust, B. T. and Gatti, H. Frandini and {Garcia-Gamez}, D. and Gariazzo, S. and Gehrlein, J. and Grant, C. and Gomes, R. A. and Hansell, A. B. and Halzen, F. and Ho, S. and Zink, J. Hoefken and Jones, R. S. and Kunkle, P. and Li, J.-Y. and Li, S. C. and Luo, X. and Malyshkin, Yu and Massaro, D. and Mastbaum, A. and Mohanta, R. and Mumm, H. P. and {Nebot-Guinot}, M. and Neilson, R. and Ni, K. and Nieves, J. and Gann, G. D. Orebi and Pandey, V. and Pascoli, S. and Qian, X. and Rajaoalisoa, M. and Roca, C. and Roskovec, B. and {Saul-Sala}, E. and Salda{\~n}a, L. and Scholberg, K. and Shakya, B. and Slocum, P. L. and Snider, E. L. and Steiger, H. Th J. and Steklain, A. F. and Stock, M. R. and Sutanto, F. and Takhistov, V. and Tsai, Y.-D. and Tsai, Y.-T. and {Venegas-Vargas}, D. and Wallbank, M. and Wang, E. and Weatherly, P. and Westerdale, S. and Worcester, E. and Wu, W. and Yang, G. and Zamorano, B.},
  year = {2023},
  number = {arXiv:2203.07323},
  eprint = {2203.07323},
  primaryclass = {astro-ph, physics:hep-ex, physics:hep-ph, physics:physics},
  publisher = {arXiv},
  doi = {10.48550/arXiv.2203.07323},
  urldate = {2024-02-09},
  archiveprefix = {arXiv},
}

@article{adrian-martinezLetterIntentKM3NeT2016,
  title = {Letter of Intent for {{KM3NeT}} 2.0},
  collaboration = "KM3NeT",
  year = {2016},
  journal = {J. Phys. G},
  volume = {43},
  number = {8},
  pages = {084001},
  publisher = {IOP Publishing},
  issn = {0954-3899},
  doi = {10.1088/0954-3899/43/8/084001},
  urldate = {2022-04-14},
}

@article{agostinelliGeant4aSimulationToolkit2003,
  title = {Geant4---a Simulation Toolkit},
  author = {Agostinelli, S. and Allison, J. and Amako, K. and Apostolakis, J. and Araujo, H. and Arce, P. and Asai, M. and Axen, D. and Banerjee, S. and Barrand, G. and Behner, F. and Bellagamba, L. and Boudreau, J. and Broglia, L. and Brunengo, A. and Burkhardt, H. and Chauvie, S. and Chuma, J. and Chytracek, R. and Cooperman, G. and Cosmo, G. and Degtyarenko, P. and Dell'Acqua, A. and Depaola, G. and Dietrich, D. and Enami, R. and Feliciello, A. and Ferguson, C. and Fesefeldt, H. and Folger, G. and Foppiano, F. and Forti, A. and Garelli, S. and Giani, S. and Giannitrapani, R. and Gibin, D. and G{\'o}mez Cadenas, J. J. and Gonz{\'a}lez, I. and Gracia Abril, G. and Greeniaus, G. and Greiner, W. and Grichine, V. and Grossheim, A. and Guatelli, S. and Gumplinger, P. and Hamatsu, R. and Hashimoto, K. and Hasui, H. and Heikkinen, A. and Howard, A. and Ivanchenko, V. and Johnson, A. and Jones, F. W. and Kallenbach, J. and Kanaya, N. and Kawabata, M. and Kawabata, Y. and Kawaguti, M. and Kelner, S. and Kent, P. and Kimura, A. and Kodama, T. and Kokoulin, R. and Kossov, M. and Kurashige, H. and Lamanna, E. and Lamp{\'e}n, T. and Lara, V. and Lefebure, V. and Lei, F. and Liendl, M. and Lockman, W. and Longo, F. and Magni, S. and Maire, M. and Medernach, E. and Minamimoto, K. and {Mora de Freitas}, P. and Morita, Y. and Murakami, K. and Nagamatu, M. and Nartallo, R. and Nieminen, P. and Nishimura, T. and Ohtsubo, K. and Okamura, M. and O'Neale, S. and Oohata, Y. and Paech, K. and Perl, J. and Pfeiffer, A. and Pia, M. G. and Ranjard, F. and Rybin, A. and Sadilov, S. and Di Salvo, E. and Santin, G. and Sasaki, T. and Savvas, N. and Sawada, Y. and Scherer, S. and Sei, S. and Sirotenko, V. and Smith, D. and Starkov, N. and Stoecker, H. and Sulkimo, J. and Takahata, M. and Tanaka, S. and Tcherniaev, E. and Safai Tehrani, E. and Tropeano, M. and Truscott, P. and Uno, H. and Urban, L. and Urban, P. and Verderi, M. and Walkden, A. and Wander, W. and Weber, H. and Wellisch, J. P. and Wenaus, T. and Williams, D. C. and Wright, D. and Yamada, T. and Yoshida, H. and Zschiesche, D.},
  year = {2003},
  journal = {Nuclear Instruments and Methods in Physics Research Section A: Accelerators, Spectrometers, Detectors and Associated Equipment},
  volume = {506},
  number = {3},
  pages = {250--303},
  issn = {0168-9002},
  doi = {10.1016/S0168-9002(03)01368-8},
  urldate = {2024-06-16},
}

@article{aguzziInertialBioluminescenceRhythms2017,
  title = {Inertial Bioluminescence Rhythms at the {{Capo Passero}} ({{KM3NeT-Italia}}) Site, {{Central Mediterranean Sea}}},
  author = {Aguzzi, J. and Fanelli, E. and Ciuffardi, T. and Schirone, A. and Craig, J.},
  year = {2017},
  journal = {Sci Rep},
  volume = {7},
  number = {1},
  pages = {44938},
  publisher = {Nature Publishing Group},
  issn = {2045-2322},
  doi = {10.1038/srep44938},
  urldate = {2023-09-12},
}

@article{aielloCharacterisationHamamatsuPhotomultipliers2018,
  title = {Characterisation of the {{Hamamatsu}} Photomultipliers for the {{KM3NeT Neutrino Telescope}}},
  collaboration = "KM3NeT",
  year = {2018},
  journal = {J. Inst.},
  volume = {13},
  number = {05},
  pages = {P05035--P05035},
  publisher = {IOP Publishing},
  issn = {1748-0221},
  doi = {10.1088/1748-0221/13/05/P05035},
  urldate = {2022-04-14},
}

@article{aielloDeterminingNeutrinoMass2022,
  title = {Determining the Neutrino Mass Ordering and Oscillation Parameters with {{KM3NeT}}/{{ORCA}}},
  collaboration = "KM3NeT",
  year = {2022},
  journal = {Eur. Phys. J. C},
  volume = {82},
  number = {1},
  pages = {26},
  issn = {1434-6052},
  doi = {10.1140/epjc/s10052-021-09893-0},
  urldate = {2024-05-15},
}

@article{aielloGSeaGenKM3NeTGENIEbased2020,
  title = {{{gSeaGen}}: {{The KM3NeT GENIE-based}} Code for Neutrino Telescopes},
  shorttitle = {{{gSeaGen}}},
  collaboration = "KM3NeT",
  year = {2020},
  journal = {Computer Physics Communications},
  volume = {256},
  pages = {107477},
  issn = {0010-4655},
  doi = {10.1016/j.cpc.2020.107477},
  urldate = {2022-10-06},
}

@article{aielloMeasurementNeutrinoOscillation2024,
  collaboration = "KM3NeT",
  title = {Measurement of Neutrino Oscillation Parameters with the First Six Detection Units of {{KM3NeT}}/{{ORCA}}},
  year = {2024},
  journal = {J. High Energ. Phys.},
  volume = {2024},
  number = {10},
  pages = {206},
  issn = {1029-8479},
  doi = {10.1007/JHEP10(2024)206},
  urldate = {2024-12-03},
}

@article{aielloNanobeaconTimeCalibration2022,
  title = {Nanobeacon: {{A}} Time Calibration Device for the {{KM3NeT}} Neutrino Telescope},
  shorttitle = {Nanobeacon},
  collaboration = "KM3NeT",
  year = {2022},
  journal = {Nuclear Instruments and Methods in Physics Research Section A: Accelerators, Spectrometers, Detectors and Associated Equipment},
  volume = {1040},
  primaryclass = {astro-ph, physics:physics},
  pages = {167132},
  issn = {01689002},
  doi = {10.1016/j.nima.2022.167132},
  urldate = {2023-06-21},
  archiveprefix = {arXiv},
}

@article{aielloProbingInvisibleNeutrino2025a,
  collaboration = "KM3NeT",
  title = {Probing Invisible Neutrino Decay with the First Six Detection Units of {{KM3NeT}}/{{ORCA}}},
  year = {2025},
  journal = {J. High Energ. Phys.},
  volume = {2025},
  number = {4},
  pages = {105},
  issn = {1029-8479},
  doi = {10.1007/JHEP04(2025)105},
  urldate = {2025-06-16},
}

@article{aielloSearchNonstandardNeutrino2025a,
  collaboration = "KM3NeT",
  title = {Search for Non-Standard Neutrino Interactions with the First Six Detection Units of {{KM3NeT}}/{{ORCA}}},
  year = {2025},
  journal = {J. Cosmol. Astropart. Phys.},
  volume = {2025},
  number = {02},
  pages = {073},
  publisher = {IOP Publishing},
  issn = {1475-7516},
  doi = {10.1088/1475-7516/2025/02/073},
  urldate = {2025-06-16},
}

@article{aielloSearchQuantumDecoherence2025,
  title = {Search for Quantum Decoherence in Neutrino Oscillations with Six Detection Units of {{KM3NeT}}/{{ORCA}}},
  collaboration = "KM3NeT",
  year = {2025},
  journal = {J. Cosmol. Astropart. Phys.},
  volume = {2025},
  number = {03},
  pages = {039},
  publisher = {IOP Publishing},
  issn = {1475-7516},
  doi = {10.1088/1475-7516/2025/03/039},
  urldate = {2025-06-16},
}

@article{aielloSensitivityLightSterile2021,
  title = {Sensitivity to Light Sterile Neutrino Mixing Parameters with {{KM3NeT}}/{{ORCA}}},
  collaboration = "KM3NeT",
  year = {2021},
  journal = {J. High Energ. Phys.},
  volume = {2021},
  number = {10},
  pages = {180},
  issn = {1029-8479},
  doi = {10.1007/JHEP10(2021)180},
  urldate = {2022-11-09},
}

@article{albertMeasuringAtmosphericNeutrino2019,
  title = {Measuring the Atmospheric Neutrino Oscillation Parameters and Constraining the 3+1 Neutrino Model with Ten Years of {{ANTARES}} Data},
  collaboration = "ANTARES",
  year = {2019},
  journal = {J. High Energ. Phys.},
  volume = {2019},
  number = {6},
  pages = {113},
  issn = {1029-8479},
  doi = {10.1007/JHEP06(2019)113},
  urldate = {2022-11-18},
}

@article{andreopoulosGENIENeutrinoMonte2010,
  title = {The {{GENIE}} Neutrino {{Monte Carlo}} Generator},
  author = {Andreopoulos, C. and Bell, A. and Bhattacharya, D. and Cavanna, F. and Dobson, J. and Dytman, S. and Gallagher, H. and Guzowski, P. and Hatcher, R. and Kehayias, P. and Meregaglia, A. and Naples, D. and Pearce, G. and Rubbia, A. and Whalley, M. and Yang, T.},
  year = {2010},
  journal = {Nuclear Instruments and Methods in Physics Research Section A: Accelerators, Spectrometers, Detectors and Associated Equipment},
  volume = {614},
  number = {1},
  pages = {87--104},
  issn = {0168-9002},
  doi = {10.1016/j.nima.2009.12.009},
  urldate = {2024-06-12},
}

@phdthesis{bailly-salinsAtmosphericMuonStudies2024,
  title = {Atmospheric Muon Studies and Light Sterile Neutrino Search with {{KM3NeT}}/{{ORCA}}},
  author = {{Bailly-Salins}, Louis},
  year = {2024},
  urldate = {2025-03-05},
  school = {Normandie Universit{\'e}},
  url = {https://theses.hal.science/tel-04823223v1},
}

@incollection{bailly-salinsSensitivityKM3NeTORCA62024,
  title = {Sensitivity of {{KM3NeT}}/{{ORCA6}} to Light Sterile Neutrino Mixing Parameters},
  collaboration = "KM3NeT",
  booktitle = {Neutrino 2024},
  author = {{Bailly-Salins}, Louis},
  year = {2024},
  address = {Milano, Italy},
  publisher = {Zenodo},
  doi = {10.5281/zenodo.13254123},
  urldate = {2024-12-03},
}

@inproceedings{bailly-salinsTimePositionOrientation2023,
  title = {Time, Position and Orientation Calibration Using Atmospheric Muons in {{KM3NeT}}},
  collaboration = "KM3NeT",
  booktitle = {{{PoS}}({{ICRC2023}})},
  author = {{Bailly-Salins}, Louis},
  year = {2023},
  volume = {444},
  pages = {218},
  publisher = {SISSA Medialab},
  doi = {10.22323/1.444.0218},
  urldate = {2023-11-09},
}

@article{barlowFittingUsingFinite1993,
  title = {Fitting Using Finite {{Monte Carlo}} Samples},
  author = {Barlow, Roger and Beeston, Christine},
  year = {1993},
  journal = {Computer Physics Communications},
  volume = {77},
  number = {2},
  pages = {219--228},
  issn = {0010-4655},
  doi = {10.1016/0010-4655(93)90005-W},
  urldate = {2024-08-23},
}

@phdthesis{bourretNeutrinoOscillationsEarth2018,
  title = {Neutrino Oscillations and Earth Tomography with {{KM3NeT-ORCA}}},
  author = {Bourret, Simon},
  year = {2018},
  urldate = {2024-09-15},
  school = {Universit{\'e} Sorbonne Paris Cit{\'e}},
  url = {https://theses.hal.science/tel-02491394v1}
}

@article{carminatiAtmosphericMUonsPArametric2008,
  title = {Atmospheric {{MUons}} from {{PArametric}} Formulas: A Fast {{GEnerator}} for Neutrino Telescopes ({{MUPAGE}})},
  shorttitle = {Atmospheric {{MUons}} from {{PArametric}} Formulas},
  author = {Carminati, G. and Bazzotti, M. and Margiotta, A. and Spurio, M.},
  year = {2008},
  journal = {Comp. Phys. Comm.},
  volume = {179},
  number = {12},
  pages = {915--923},
  issn = {0010-4655},
  doi = {10.1016/j.cpc.2008.07.014},
  urldate = {2023-06-20},
}

@phdthesis{carreterocuencaNeutrinoOscillationsInvisible2024,
  title = {Neutrino Oscillations and Invisible Decay with the {{KM3NeT}}/{{ORCA}} Detector},
  author = {Carretero Cuenca, V{\'i}ctor},
  year = {2024},
  address = {Valencia},
  school = {Universitat de Valencia},
  url = {https://hdl.handle.net/10550/96472},
  urldate = {2024-04-10},
}

@phdthesis{chauStudyAtmosphericNeutrino2021,
  title = {Study of Atmospheric Neutrino Oscillations with the Deep-Sea {{Cherenkov}} Detector {{KM3NeT}}/{{ORCA}} and Synergies with Reactor Neutrinos},
  author = {Chau, Thien Nhan},
  year = {2021},
  urldate = {2024-08-22},
  school = {Universit{\'e} Paris Cit{\'e}},
  url = {https://theses.hal.science/tel-03999509}
}

@article{chiarusiKM3NeTDataAcquisition2023,
  title = {The {{KM3NeT}} Data Acquisition System - {{Status}} and Evolution},
  author = {Chiarusi, Tommaso and Giorgio, Emidio and Zito, Daniele},
  collaboration = "KM3NeT",
  year = {2023},
  journal = {EPJ Web Conf.},
  volume = {280},
  pages = {08004},
  publisher = {EDP Sciences},
  issn = {2100-014X},
  doi = {10.1051/epjconf/202328008004},
  urldate = {2025-03-06},
}

@software{joao_coelho_2023_10104847,
  author       = {Joao Coelho and
                  Rebekah Pestes and
                  Alba Domi and
                  Simon Bourret and
                  Ushak Rahaman and
                  Lukas Maderer and
                  Víctor Carretero},
  title        = {joaoabcoelho/OscProb: v2.0.12},
  year         = 2023,
  publisher    = {Zenodo},
  version      = {v2.0.12},
  doi          = {10.5281/zenodo.10104847},
  url          = {https://doi.org/10.5281/zenodo.10104847}
}

@incollection{conwayIncorporatingNuisanceParameters2011,
  title = {Incorporating {{Nuisance Parameters}} in {{Likelihoods}} for {{Multisource Spectra}}},
  author = {Conway, J.S.},
  booktitle = {the  proceedings of the {PHYSTAT 2011}},
  year = {2011},
  pages = {115--120},
  doi = {10.5170/CERN-2011-006.115},
  urldate = {2024-08-23},
}

@article{dayabaycollaborationPrecisionMeasurementReactor2023,
  title = {Precision {{Measurement}} of {{Reactor Antineutrino Oscillation}} at {{Kilometer-Scale Baselines}} by {{Daya Bay}}},
  collaboration = "Daya Bay",
  year = {2023},
  journal = {Phys. Rev. Lett.},
  volume = {130},
  number = {16},
  pages = {161802},
  publisher = {American Physical Society},
  doi = {10.1103/PhysRevLett.130.161802},
  urldate = {2024-05-21},
}

@techreport{dentonSnowmassNeutrinoFrontier2022,
  title = {Snowmass {{Neutrino Frontier}}: {{NF01 Topical Group Report}} on {{Three-Flavor Neutrino Oscillations}}},
  shorttitle = {Snowmass {{Neutrino Frontier}}},
  author = {Denton, Peter B. and Friend, Megan and Messier, Mark D. and Tanaka, Hirohisa A. and B{\"o}ser, Sebastian and Coelho, Jo{\~a}o A. B. and {Perrin-Terrin}, Mathieu and Stuttard, Tom},
  year = {2022},
  number = {arXiv:2212.00809},
  eprint = {2212.00809},
  primaryclass = {hep-ex, physics:hep-ph},
  publisher = {arXiv},
  doi = {10.48550/arXiv.2212.00809},
  urldate = {2024-08-11},
  archiveprefix = {arXiv},
}

@phdthesis{domiShowerReconstructionSterile2019a,
  type = {These de Doctorat},
  title = {Shower Reconstruction and Sterile Neutrino Analysis with {{KM3NeT}}/{{ORCA}} and {{Antares}}},
  author = {Domi, Alba},
  year = {2019},
  urldate = {2024-05-23},
  school = {Aix-Marseille},
  url = {https://theses.fr/2019AIXM0550},
}

@article{dziewonskiPreliminaryReferenceEarth1981,
  title = {Preliminary Reference {{Earth}} Model},
  author = {Dziewonski, Adam M. and Anderson, Don L.},
  year = {1981},
  journal = {Physics of the Earth and Planetary Interiors},
  volume = {25},
  number = {4},
  pages = {297--356},
  issn = {0031-9201},
  doi = {10.1016/0031-9201(81)90046-7},
  urldate = {2024-08-20},
}

@article{ermilovav.k.BuildupNeutrinoOscillations1986,
  title = {Buildup of Neutrino Oscillations in the {{Earth}}},
  author = {{Ermilova, V. K.} and {Tsarev, V. A.} and {Chechin, V. A.}},
  year = {1986},
  journal = {JETP Letters},
  volume = {43},
  number = {8},
  pages = {353},
  urldate = {2024-08-16},
}

@article{esmailiConstrainingSterileNeutrinos2012,
  title = {Constraining Sterile Neutrinos with {{AMANDA}} and {{IceCube}} Atmospheric Neutrino Data},
  author = {Esmaili, Arman and Halzen, Francis and Peres, O. L. G.},
  year = {2012},
  journal = {J. Cosmol. Astropart. Phys.},
  volume = {2012},
  number = {11},
  pages = {041},
  issn = {1475-7516},
  doi = {10.1088/1475-7516/2012/11/041},
  urldate = {2024-08-19},
}

@article{estebanFateHintsUpdated2020,
  title = {The Fate of Hints: Updated Global Analysis of Three-Flavor Neutrino Oscillations},
  shorttitle = {The Fate of Hints},
  author = {Esteban, Ivan and {Gonzalez-Garcia}, M.C. and Maltoni, Michele and Schwetz, Thomas and Zhou, Albert},
  year = {2020},
  journal = {J. High Energ. Phys.},
  volume = {2020},
  number = {9},
  pages = {178},
  issn = {1029-8479},
  doi = {10.1007/JHEP09(2020)178},
  urldate = {2024-08-11},
}

@inproceedings{gatiusoliverDynamicalPositionOrientation2023,
  title = {Dynamical Position and Orientation Calibration of the {{KM3NeT}} Telescope},
  collaboration = "KM3NeT",
  booktitle = {{{PoS}}({{ICRC2023}})},
  author = {Gatius Oliver, Clara and Bretaudeau, F{\'e}lix and De Jong, Maarten and Martin, Lilian},
  year = {2023},
  volume = {444},
  pages = {1033},
  publisher = {SISSA Medialab},
  doi = {10.22323/1.444.1033},
  urldate = {2023-10-04},
}

@book{giuntiFundamentalsNeutrinoPhysics2007,
  title = {Fundamentals of {{Neutrino Physics}} and {{Astrophysics}}},
  author = {Giunti, Carlo and Kim, Chung Wook},
  year = {2007},
  publisher = {Oxford University Press, UK},
  address = {Oxford},
  doi = {10.1093/acprof:oso/9780198508717.001.0001},
}

@article{haddockBioluminescenceSea2010,
  title = {Bioluminescence in the Sea},
  author = {Haddock, Steven H. D. and Moline, Mark A. and Case, James F.},
  year = {2010},
  journal = {Ann Rev Mar Sci},
  volume = {2},
  pages = {443--493},
  issn = {1941-1405},
  doi = {10.1146/annurev-marine-120308-081028},
}

@article{hondaAtmosphericNeutrinoFlux2015,
  title = {Atmospheric Neutrino Flux Calculation Using the {{NRLMSISE-00}} Atmospheric Model},
  author = {Honda, M. and Athar, M. Sajjad and Kajita, T. and Kasahara, K. and Midorikawa, S.},
  year = {2015},
  journal = {Phys. Rev. D},
  volume = {92},
  number = {2},
  pages = {023004},
  publisher = {American Physical Society},
  doi = {10.1103/PhysRevD.92.023004},
  urldate = {2024-08-13},
}

@article{km3netcollaborationDependenceAtmosphericMuon2020,
  title = {Dependence of Atmospheric Muon Flux on Seawater Depth Measured with the First {{KM3NeT}} Detection Units},
  collaboration = "KM3NeT",
  year = {2020},
  journal = {Eur. Phys. J. C},
  volume = {80},
  number = {2},
  pages = {99},
  issn = {1434-6044, 1434-6052},
  doi = {10.1140/epjc/s10052-020-7629-z},
  urldate = {2022-04-14},
  archiveprefix = {arXiv},
}

@article{mikheyevResonantNeutrinoOscillations1989,
  title = {Resonant Neutrino Oscillations in Matter},
  author = {Mikheyev, S. P. and Smirnov, A. {Yu}.},
  year = {1989},
  journal = {Progress in Particle and Nuclear Physics},
  volume = {23},
  pages = {41--136},
  issn = {0146-6410},
  doi = {10.1016/0146-6410(89)90008-2},
  urldate = {2024-08-16},
}

@article{nunokawaProbingLSNDMass2003,
  title = {Probing the {{LSND}} Mass Scale and Four Neutrino Scenarios with a Neutrino Telescope},
  author = {Nunokawa, H. and Peres, O. L. G. and Zukanovich Funchal, R.},
  year = {2003},
  journal = {Physics Letters B},
  volume = {562},
  number = {3},
  pages = {279--290},
  issn = {0370-2693},
  doi = {10.1016/S0370-2693(03)00603-8},
  urldate = {2024-08-19},
}

@phdthesis{ofearraighFollowingLightNovel2024,
  title = {Following the Light - {{Novel}} Event Reconstruction Techniques for Neutrino Oscillation Analyses in {{KM3NeT}}/{{ORCA}}},
  author = {O Fearraigh, Brian},
  year = {2024},
  school = {Amsterdam University},
  url = {https://dare.uva.nl/search?identifier=c0d5bde5-b513-4291-a4df-fbf4d5c69add}
}

@article{PrecisionElectroweakMeasurements2006,
  title = {Precision Electroweak Measurements on the {{Z}} Resonance},
  year = {2006},
  collaborations = "ALEPH, DELPHI, L3, OPAL, SLD",
  journal = {Physics Reports},
  volume = {427},
  number = {5},
  pages = {257--454},
  issn = {0370-1573},
  doi = {10.1016/j.physrep.2005.12.006},
  urldate = {2024-04-09},
}

@misc{ROOTMinuit2MnMigrad,
  title = {{{ROOT}}::{{Minuit2}}::{{MnMigrad Class Reference}}},
  urldate = {2024-08-23},
  howpublished = {https://root.cern.ch/doc/master/classROOT\_1\_1Minuit2\_1\_1MnMigrad.html}
}

@inproceedings{serranoWhiteRabbitProject2009,
  title = {The {{White Rabbit Project}}},
  booktitle = {Proceedings of {{ICALEPCS TUC004}}},
  author = {Serrano, J and Alvarez, P and Cattin, M and Cota, E. G. and Lewis, P. M. J. H.},
  year = {2009},
  address = {Kobe, Japan},
  url = {https://proceedings.jacow.org/icalepcs2009/papers/tuc004.pdf}
}

@article{super-kamiokandecollaborationEvidenceOscillationAtmospheric1998,
  title = {Evidence for {{Oscillation}} of {{Atmospheric Neutrinos}}},
  collaboration = "Super-Kamiokande",
  year = {1998},
  journal = {Phys. Rev. Lett.},
  volume = {81},
  number = {8},
  pages = {1562--1567},
  publisher = {American Physical Society},
  doi = {10.1103/PhysRevLett.81.1562},
  urldate = {2024-08-06}
}

@article{super-kamiokandecollaborationLimitsSterileNeutrino2015,
  title = {Limits on Sterile Neutrino Mixing Using Atmospheric Neutrinos in {{Super-Kamiokande}}},
  collaboration = "Super-Kamiokande",
  year = {2015},
  journal = {Phys. Rev. D},
  volume = {91},
  number = {5},
  pages = {052019},
  publisher = {American Physical Society},
  doi = {10.1103/PhysRevD.91.052019},
  urldate = {2024-04-15},
}

@article{thenovacollaborationSearchActiveSterileAntineutrino2021,
  title = {Search for {{Active-Sterile Antineutrino Mixing Using Neutral-Current Interactions}} with the {{NOvA Experiment}}},
  collaboration = "NOvA",
  year = {2021},
  journal = {Phys. Rev. Lett.},
  volume = {127},
  number = {20},
  pages = {201801},
  publisher = {American Physical Society},
  doi = {10.1103/PhysRevLett.127.201801},
  urldate = {2024-08-12},
}

@article{tsirigotisHOUReconstructionSimulation2011,
  title = {{{HOU Reconstruction}} \& {{Simulation}} ({{HOURS}}): {{A}} Complete Simulation and Reconstruction Package for Very Large Volume Underwater Neutrino Telescopes},
  shorttitle = {{{HOU Reconstruction}} \& {{Simulation}} ({{HOURS}})},
  author = {Tsirigotis, A. G. and Leisos, A. and Tzamarias, S. E.},
  year = {2011},
  journal = {Nuclear Instruments and Methods in Physics Research Section A: Accelerators, Spectrometers, Detectors and Associated Equipment},
  volume = {626--627},
  pages = {S185-S187},
  issn = {0168-9002},
  doi = {10.1016/j.nima.2010.06.258},
  urldate = {2025-03-28},
}

@article{widderBioluminescencePelagicVisual2002a,
  title = {Bioluminescence and the {{Pelagic Visual Environment}}},
  author = {Widder, Edith},
  year = {2002},
  journal = {Marine and Freshwater Behaviour and Physiology},
  volume = {35},
  number = {1-2},
  pages = {1--26},
  publisher = {Taylor \& Francis},
  issn = {1023-6244},
  doi = {10.1080/10236240290025581},
  urldate = {2023-09-12},
}

@article{wolfensteinNeutrinoOscillationsMatter1978,
  title = {Neutrino Oscillations in Matter},
  author = {Wolfenstein, L.},
  year = {1978},
  journal = {Phys. Rev. D},
  volume = {17},
  number = {9},
  pages = {2369--2374},
  publisher = {American Physical Society},
  doi = {10.1103/PhysRevD.17.2369},
  urldate = {2024-08-16},
}

@article{lesgourguesMassiveNeutrinosCosmology2006,
  title = {Massive Neutrinos and Cosmology},
  author = {Lesgourgues, Julien and Pastor, Sergio},
  year = {2006},
  journal = {Physics Reports},
  volume = {429},
  number = {6},
  pages = {307--379},
  issn = {0370-1573},
  doi = {10.1016/j.physrep.2006.04.001},
  urldate = {2025-07-21},
}

@article{abbasiExplorationMassSplitting2024a,
  title = {Exploration of Mass Splitting and Muon/Tau Mixing Parameters for an {{eV-scale}} Sterile Neutrino with {{IceCube}}},
  collaboration = "IceCube",
  year = {2024},
  journal = {Physics Letters B},
  volume = {858},
  pages = {139077},
  issn = {0370-2693},
  doi = {10.1016/j.physletb.2024.139077},
  urldate = {2025-07-21},
}

@article{aielloAtmosphericMuonsMeasured2024,
  title = {Atmospheric Muons Measured with the {{KM3NeT}} Detectors in Comparison with Updated Numeric Predictions},
  collaboration = "KM3NeT",
  year = {2024},
  journal = {Eur. Phys. J. C},
  volume = {84},
  number = {7},
  pages = {696},
  issn = {1434-6052},
  doi = {10.1140/epjc/s10052-024-13018-8},
  urldate = {2024-08-14},
}

@article{icecubecollaborationSearchEVScaleSterile2024a,
  title = {Search for an {{eV-Scale Sterile Neutrino Using Improved High-Energy}} $\nu_\mu$ {{Event Reconstruction}} in {{IceCube}}},
  collaboration = "IceCube",
  year = {2024},
  journal = {Phys. Rev. Lett.},
  volume = {133},
  number = {20},
  pages = {201804},
  publisher = {American Physical Society},
  doi = {10.1103/PhysRevLett.133.201804},
  urldate = {2025-07-21},
}

@article{icecubecollaborationSearchLightSterile2024,
  title = {Search for a Light Sterile Neutrino with 7.5 Years of {{IceCube DeepCore}} Data},
  collaboration = "IceCube",
  year = {2024},
  journal = {Phys. Rev. D},
  volume = {110},
  number = {7},
  pages = {072007},
  publisher = {American Physical Society},
  doi = {10.1103/PhysRevD.110.072007},
  urldate = {2025-07-21},
}

@article{aielloKM3NeTMultiPMTOptical2022a,
  title = {The {{KM3NeT}} Multi-{{PMT}} Optical Module},
  collaboration = "KM3NeT",
  year = {2022},
  journal = {J. Inst.},
  volume = {17},
  number = {07},
  pages = {P07038},
  publisher = {IOP Publishing},
  issn = {1748-0221},
  doi = {10.1088/1748-0221/17/07/P07038},
  urldate = {2023-09-08},
}

@article{aielloKM3NeTBroadcastOptical2023,
  title = {{{KM3NeT}} Broadcast Optical Data Transport System},
  collaboration = "KM3NeT",
  year = {2023},
  journal = {J. Inst.},
  volume = {18},
  number = {02},
  pages = {T02001},
  publisher = {IOP Publishing},
  issn = {1748-0221},
  doi = {10.1088/1748-0221/18/02/T02001},
  urldate = {2023-09-08},
}

@article{ambrosioMeasurementAtmosphericNeutrinoinduced1998,
  title = {Measurement of the Atmospheric Neutrino-Induced Upgoing Muon Flux Using {{MACRO}}},
  author = {Ambrosio, M and Antolini, R and Aramo, C and Auriemma, G and Baldini, A and C. Barbarino, G and C. Barish, B and Battistoni, G and Bellotti, R and Bemporad, C and Bernardini, P and Bilokon, H and Bisi, V and Bloise, C and Bower, C and Bussino, S and Cafagna, F and Calicchio, M and Campana, D and Carboni, M and Castellano, M and Cecchini, S and Cei, F and Chiarella, V and Choudhary, B. C and Coutu, S and De Benedictis, L and De Cataldo, G and Dekhissi, H and De Marzo, C and De Mitri, I and Derkaoui, J and De Vincenzi, M and Di Credico, A and Erriquez, O and Favuzzi, C and Forti, C and Fusco, P and Giacomelli, G and Giannini, G and Giglietto, N and Giorgini, M and Grassi, M and Gray, L and Grillo, A and Guarino, F and Guarnaccia, P and Gustavino, C and Habig, A and Hanson, K and Hawthorne, A and Heinz, R and Huang, Y and Iarocci, E and Katsavounidis, E and Katsavounidis, I and Kearns, E and Kim, H and Kyriazopoulou, S and Lamanna, E and Lane, C and Levin, D. S and Lipari, P and Longley, N. P and Longo, M. J and Maaroufi, F and Mancarella, G and Mandrioli, G and Manzoor, S and Margiotta Neri, A and Marini, A and Martello, D and {Marzari-Chiesa}, A and Mazziotta, M. N and Mazzotta, C and Michael, D. G and Mikheyev, S and Miller, L and Monacelli, P and Montaruli, T and Monteno, M and Mufson, S and Musser, J and Nicol{\'o}, D and Nolty, R and Okada, C and Orth, C and Osteria, G and Ouchrif, M and Palamara, O and Patera, V and Patrizii, L and Pazzi, R and Peck, C. W and Petrera, S and Pistilli, P and Popa, V and Pugliese, V and Rain{\'o}, A and Reynoldson, J and Ronga, F and Rubizzo, U and Sanzgiri, A and Satriano, C and Satta, L and Scapparone, E and Scholberg, K and Sciubba, A and {Serra-Lugaresi}, P and Severi, M and Sioli, M and Sitta, M and Spinelli, P and Spinetti, M and Spurio, M and Steinberg, R and Stone, J. L and Sulak, L. R and Surdo, A and Tarl{\'e}, G and Togo, V and Ugolotti, D and Vakili, M and Walter, C. W and Webb, R},
  collaboration = "MACRO",
  year = {1998},
  journal = {Physics Letters B},
  volume = {434},
  number = {3},
  pages = {451--457},
  issn = {0370-2693},
  doi = {10.1016/S0370-2693(98)00885-5}
}

@inproceedings{melisInSituCalibrationKM3NeT2017,
  title = {In-{{Situ Calibration}} of {{KM3NeT}}},
  booktitle = {Proceedings of 35th {{International Cosmic Ray Conference}} --- {{PoS}}({{ICRC2017}})},
  author = {Melis, Karel},
  collaboration = "KM3NeT",
  year = {2017},
  pages = {1059},
  publisher = {Sissa Medialab},
  address = {Bexco, Busan, Korea},
  doi = {10.22323/1.301.1059}
}

@article{razzaqueSearchesSterileNeutrinos2012,
  title = {Searches for Sterile Neutrinos with {{IceCube DeepCore}}},
  author = {Razzaque, Soebur and Smirnov, A. {Yu}.},
  year = {2012},
  journal = {Phys. Rev. D},
  volume = {85},
  number = {9},
  pages = {093010},
  publisher = {American Physical Society},
  doi = {10.1103/PhysRevD.85.093010},
  urldate = {2024-08-19},
}

@article{razzaqueSearchingSterileNeutrinos2011,
  title = {Searching for Sterile Neutrinos in Ice},
  author = {Razzaque, Soebur and Smirnov, A. {Yu}.},
  year = {2011},
  journal = {J. High Energ. Phys.},
  volume = {2011},
  number = {7},
  pages = {84},
  issn = {1029-8479},
  doi = {10.1007/JHEP07(2011)084},
  urldate = {2024-08-20},
}

@article{aielloStudyTauNeutrinos2025,
  title = {Study of Tau Neutrinos and Non-Unitary Neutrino Mixing with the First Six Detection Units of {{KM3NeT}}/{{ORCA}}},
  author = {Aiello, S. and Albert, A. and Alhebsi, A. R. and Alshamsi, M. and Alves Garre, S. and Ambrosone, A. and Ameli, F. and Andre, M. and Aphecetche, L. and Ardid, M. and Ardid, S. and Aublin, J. and Badaracco, F. and {Bailly-Salins}, L. and Barda{\v c}ov{\'a}, Z. and Baret, B. and {Bariego-Quintana}, A. and Becherini, Y. and Bendahman, M. and Benfenati Gualandi, F. and Benhassi, M. and Bennani, M. and Benoit, D. M. and Berbee, E. and Bertin, V. and Biagi, S. and Boettcher, M. and Bonanno, D. and Bouasla, A. B. and Boumaaza, J. and Bouta, M. and Bouwhuis, M. and Bozza, C. and Bozza, R. M. and Br{\^a}nza{\c s}, H. and Bretaudeau, F. and Breuhaus, M. and Bruijn, R. and Brunner, J. and Bruno, R. and Buis, E. and Buompane, R. and Busto, J. and Caiffi, B. and Calvo, D. and Capone, A. and Carenini, F. and Carretero, V. and Cartraud, T. and Castaldi, P. and Cecchini, V. and Celli, S. and Cerisy, L. and Chabab, M. and Chen, A. and Cherubini, S. and Chiarusi, T. and Circella, M. and Clark, R. and Cocimano, R. and Coelho, J. A. B. and Coleiro, A. and Condorelli, A. and Coniglione, R. and Coyle, P. and Creusot, A. and Cuttone, G. and Dallier, R. and De Benedittis, A. and De Wasseige, G. and Decoene, V. and Deguire, P. and Del Rosso, I. and Di Mauro, L. S. and Di Palma, I. and D{\'i}az, A. F. and {Diego-Tortosa}, D. and Distefano, C. and Domi, A. and Donzaud, C. and Dornic, D. and Drakopoulou, E. and Drouhin, D. and Ducoin, J.-G. and Duverne, P. and Dvornick{\'y}, R. and Eberl, T. and Eckerov{\'a}, E. and Eddymaoui, A. and {van Eeden}, T. and Eff, M. and {van Eijk}, D. and El Bojaddaini, I. and El Hedri, S. and El Mentawi, S. and Ellajosyula, V. and Enzenh{\"o}fer, A. and Ferrara, G. and Filipovi{\'c}, M. D. and Filippini, F. and Franciotti, D. and Fusco, L. A. and Gagliardini, S. and Gal, T. and Garc{\'i}a M{\'e}ndez, J. and Garcia Soto, A. and Gatius Oliver, C. and Gei{\ss}elbrecht, N. and Genton, E. and Ghaddari, H. and Gialanella, L. and Gibson, B. K. and Giorgio, E. and Goos, I. and Goswami, P. and Gozzini, S. R. and Gracia, R. and Guidi, C. and Guillon, B. and Guti{\'e}rrez, M. and Haack, C. and {van Haren}, H. and Heijboer, A. and Hennig, L. and {Hern{\'a}ndez-Rey}, J. J. and Idrissi, A. and Idrissi Ibnsalih, W. and Illuminati, G. and Joly, D. and {de Jong}, M. and {de Jong}, P. and Jung, B. J. and Kalaczy{\'n}ski, P. and Kikvadze, V. and Kistauri, G. and Kopper, C. and Kouchner, A. and Kovalev, Y. Y. and Krupa, L. and Kueviakoe, V. and Kulikovskiy, V. and Kvatadze, R. and Labalme, M. and Lahmann, R. and Lamoureux, M. and Larosa, G. and Lastoria, C. and Lazar, J. and Lazo, A. and Le Stum, S. and Lehaut, G. and Lema{\^i}tre, V. and Leonora, E. and Lessing, N. and Levi, G. and Lindsey Clark, M. and Longhitano, F. and Magnani, F. and Majumdar, J. and Malerba, L. and Mamedov, F. and Manfreda, A. and Manousakis, A. and Marconi, M. and Margiotta, A. and Marinelli, A. and Markou, C. and Martin, L. and Mastrodicasa, M. and Mastroianni, S. and Mauro, J. and Mehta, K. C. K. and Meskar, A. and Miele, G. and Migliozzi, P. and Migneco, E. and Mitsou, M. L. and Mollo, C. M. and {Morales-Gallegos}, L. and Moussa, A. and Mozun Mateo, I. and Muller, R. and Musone, M. R. and Musumeci, M. and Navas, S. and Nayerhoda, A. and Nicolau, C. A. and Nkosi, B. and Fearraigh, B. {\'O} and Oliviero, V. and Orlando, A. and Oukacha, E. and Paesani, D. and Palacios Gonz{\'a}lez, J. and Papalashvili, G. and Parisi, V. and Parmar, A. and Pastor Gomez, E. J. and Pastore, C. and P{\u a}un, A. M. and P{\u a}v{\u a}la{\c s}, G. E. and Pe{\~n}a Mart{\'i}nez, S. and {Perrin-Terrin}, M. and Pestel, V. and Pestes, R. and Piattelli, P. and Plavin, A. and Poir{\`e}, C. and Popa, V. and Pradier, T. and Prado, J. and Pulvirenti, S. and {Quiroz-Rangel}, C. A. and Randazzo, N. and Ratnani, A. and Razzaque, S. and Rea, I. C. and Real, D. and Riccobene, G. and Romanov, A. and Ros, E. and {\v S}aina, A. and Salesa Greus, F. and Samtleben, D. F. E. and S{\'a}nchez Losa, A. and Sanfilippo, S. and Sanguineti, M. and Santonocito, D. and Sapienza, P. and Scarnera, M. and Schnabel, J. and Schumann, J. and Schutte, H. M. and Seneca, J. and Sennan, N. and Sevle, P. and Sgura, I. and Shanidze, R. and Sharma, A. and Shitov, Y. and {\v S}imkovic, F. and Simonelli, A. and Sinopoulou, A. and Spisso, B. and Spurio, M. and Stavropoulos, D. and {\v S}tekl, I. and Taiuti, M. and Takadze, G. and Tayalati, Y. and Thiersen, H. and Thoudam, S. and {Tosta e Melo}, I. and Trocm{\'e}, B. and Tsourapis, V. and Tudorache, A. and Tzamariudaki, E. and Ukleja, A. and Vacheret, A. and Valsecchi, V. and Van Elewyck, V. and Vannoye, G. and Vasileiadis, G. and {Vazquez de Sola}, F. and Veutro, A. and Viola, S. and Vivolo, D. and {van Vliet}, A. and {de Wolf}, E. and {Lhenry-Yvon}, I. and Zavatarelli, S. and Zegarelli, A. and Zito, D. and Zornoza, J. D. and Z{\'u}{\~n}iga, J. and Zywucka, N. and {The KM3NeT collaboration}},
  year = {2025},
  journal = {J. High Energ. Phys.},
  volume = {2025},
  number = {7},
  pages = {213},
  issn = {1029-8479},
  doi = {10.1007/JHEP07(2025)213},
  urldate = {2025-10-04}
}

@article{particledatagroupReviewParticlePhysics2018,
  title = {Review of {{Particle Physics}}},
  author = {{Particle Data Group} and Tanabashi, M. and Hagiwara, K. and Hikasa, K. and Nakamura, K. and Sumino, Y. and Takahashi, F. and Tanaka, J. and Agashe, K. and Aielli, G. and Amsler, C. and Antonelli, M. and Asner, D. M. and Baer, H. and Banerjee, {\relax Sw}. and Barnett, R. M. and Basaglia, T. and Bauer, C. W. and Beatty, J. J. and Belousov, V. I. and Beringer, J. and Bethke, S. and Bettini, A. and Bichsel, H. and Biebel, O. and Black, K. M. and Blucher, E. and Buchmuller, O. and Burkert, V. and Bychkov, M. A. and Cahn, R. N. and Carena, M. and Ceccucci, A. and Cerri, A. and Chakraborty, D. and Chen, M.-C. and Chivukula, R. S. and Cowan, G. and Dahl, O. and D'Ambrosio, G. and Damour, T. and {de Florian}, D. and {de Gouv{\^e}a}, A. and DeGrand, T. and {de Jong}, P. and Dissertori, G. and Dobrescu, B. A. and D'Onofrio, M. and Doser, M. and Drees, M. and Dreiner, H. K. and Dwyer, D. A. and Eerola, P. and Eidelman, S. and Ellis, J. and Erler, J. and Ezhela, V. V. and Fetscher, W. and Fields, B. D. and Firestone, R. and Foster, B. and Freitas, A. and Gallagher, H. and Garren, L. and Gerber, H.-J. and Gerbier, G. and Gershon, T. and Gershtein, Y. and Gherghetta, T. and Godizov, A. A. and Goodman, M. and Grab, C. and Gritsan, A. V. and Grojean, C. and Groom, D. E. and Gr{\"u}newald, M. and Gurtu, A. and Gutsche, T. and Haber, H. E. and Hanhart, C. and Hashimoto, S. and Hayato, Y. and Hayes, K. G. and Hebecker, A. and Heinemeyer, S. and Heltsley, B. and {Hern{\'a}ndez-Rey}, J. J. and Hisano, J. and H{\"o}cker, A. and Holder, J. and Holtkamp, A. and Hyodo, T. and Irwin, K. D. and Johnson, K. F. and Kado, M. and Karliner, M. and Katz, U. F. and Klein, S. R. and Klempt, E. and Kowalewski, R. V. and Krauss, F. and Kreps, M. and Krusche, B. and Kuyanov, {\relax Yu}. V. and Kwon, Y. and Lahav, O. and Laiho, J. and Lesgourgues, J. and Liddle, A. and Ligeti, Z. and Lin, C.-J. and Lippmann, C. and Liss, T. M. and Littenberg, L. and Lugovsky, K. S. and Lugovsky, S. B. and Lusiani, A. and Makida, Y. and Maltoni, F. and Mannel, T. and Manohar, A. V. and Marciano, W. J. and Martin, A. D. and Masoni, A. and Matthews, J. and Mei{\ss}ner, U.-G. and Milstead, D. and Mitchell, R. E. and M{\"o}nig, K. and Molaro, P. and Moortgat, F. and Moskovic, M. and Murayama, H. and Narain, M. and Nason, P. and Navas, S. and Neubert, M. and Nevski, P. and Nir, Y. and Olive, K. A. and Pagan Griso, S. and Parsons, J. and Patrignani, C. and Peacock, J. A. and Pennington, M. and Petcov, S. T. and Petrov, V. A. and Pianori, E. and Piepke, A. and Pomarol, A. and Quadt, A. and Rademacker, J. and Raffelt, G. and Ratcliff, B. N. and Richardson, P. and Ringwald, A. and Roesler, S. and Rolli, S. and Romaniouk, A. and Rosenberg, L. J. and Rosner, J. L. and Rybka, G. and Ryutin, R. A. and Sachrajda, C. T. and Sakai, Y. and Salam, G. P. and Sarkar, S. and Sauli, F. and Schneider, O. and Scholberg, K. and Schwartz, A. J. and Scott, D. and Sharma, V. and Sharpe, S. R. and Shutt, T. and Silari, M. and Sj{\"o}strand, T. and Skands, P. and Skwarnicki, T. and Smith, J. G. and Smoot, G. F. and Spanier, S. and Spieler, H. and Spiering, C. and Stahl, A. and Stone, S. L. and Sumiyoshi, T. and Syphers, M. J. and Terashi, K. and Terning, J. and Thoma, U. and Thorne, R. S. and Tiator, L. and Titov, M. and Tkachenko, N. P. and T{\"o}rnqvist, N. A. and Tovey, D. R. and Valencia, G. and {Van de Water}, R. and Varelas, N. and Venanzoni, G. and Verde, L. and Vincter, M. G. and Vogel, P. and Vogt, A. and Wakely, S. P. and Walkowiak, W. and Walter, C. W. and Wands, D. and Ward, D. R. and Wascko, M. O. and Weiglein, G. and Weinberg, D. H. and Weinberg, E. J. and White, M. and Wiencke, L. R. and Willocq, S. and Wohl, C. G. and Womersley, J. and Woody, C. L. and Workman, R. L. and Yao, W.-M. and Zeller, G. P. and Zenin, O. V. and Zhu, R.-Y. and Zhu, S.-L. and Zimmermann, F. and Zyla, P. A. and Anderson, J. and Fuller, L. and Lugovsky, V. S. and Schaffner, P.},
  collaboration = {{Particle Data Group}},
  year = 2018,
  journal = {Phys. Rev. D},
  volume = {98},
  number = {3},
  pages = {030001},
  publisher = {American Physical Society},
  doi = {10.1103/PhysRevD.98.030001},
  urldate = {2024-08-13},
}

@article{particledatagroupReviewParticlePhysics2024,
  title = {Review of {{Particle Physics}}},
  author = {{Particle Data Group} and Navas, S. and Amsler, C. and Gutsche, T. and Hanhart, C. and {Hern{\'a}ndez-Rey}, J. J. and Louren{\c c}o, C. and Masoni, A. and Mikhasenko, M. and Mitchell, R. E. and Patrignani, C. and Schwanda, C. and Spanier, S. and Venanzoni, G. and Yuan, C. Z. and Agashe, K. and Aielli, G. and Allanach, B. C. and {Alvarez-Mu{\~n}iz}, J. and Antonelli, M. and Aschenauer, E. C. and Asner, D. M. and Assamagan, K. and Baer, H. and Banerjee, {\relax Sw}. and Barnett, R. M. and Baudis, L. and Bauer, C. W. and Beatty, J. J. and Beringer, J. and Bettini, A. and Biebel, O. and Black, K. M. and Blucher, E. and Bonventre, R. and Briere, R. A. and Buckley, A. and Burkert, V. D. and Bychkov, M. A. and Cahn, R. N. and Cao, Z. and Carena, M. and Casarosa, G. and Ceccucci, A. and Cerri, A. and Chivukula, R. S. and Cowan, G. and Cranmer, K. and Crede, V. and Cremonesi, O. and D'Ambrosio, G. and Damour, T. and {de Florian}, D. and {de Gouv{\^e}a}, A. and DeGrand, T. and Demers, S. and Demiragli, Z. and Dobrescu, B. A. and D'Onofrio, M. and Doser, M. and Dreiner, H. K. and Eerola, P. and Egede, U. and Eidelman, S. and {El-Khadra}, A. X. and Ellis, J. and Eno, S. C. and Erler, J. and Ezhela, V. V. and Fava, A. and Fetscher, W. and Fields, B. D. and Freitas, A. and Gallagher, H. and Gershon, T. and Gershtein, Y. and Gherghetta, T. and {Gonzalez-Garcia}, M. C. and Goodman, M. and Grab, C. and Gritsan, A. V. and Grojean, C. and Groom, D. E. and Gr{\"u}newald, M. and Gurtu, A. and Haber, H. E. and Hamel, M. and Hashimoto, S. and Hayato, Y. and Hebecker, A. and Heinemeyer, S. and Hikasa, K. and Hisano, J. and H{\"o}cker, A. and Holder, J. and Hsu, L. and Huston, J. and Hyodo, T. and Ianni, {\relax Al}. and Kado, M. and Karliner, M. and Katz, U. F. and Kenzie, M. and Khoze, V. A. and Klein, S. R. and Krauss, F. and Kreps, M. and Kri{\v z}an, P. and Krusche, B. and Kwon, Y. and Lahav, O. and Lellouch, L. P. and Lesgourgues, J. and Liddle, A. R. and Ligeti, Z. and Lin, C.-J. and Lippmann, C. and Liss, T. M. and Lister, A. and Littenberg, L. and Lugovsky, K. S. and Lugovsky, S. B. and Lusiani, A. and Makida, Y. and Maltoni, F. and Manohar, A. V. and Marciano, W. J. and Matthews, J. and Mei{\ss}ner, U.-G. and {Melzer-Pellmann}, I.-A. and Mertsch, P. and Miller, D. J. and Milstead, D. and M{\"o}nig, K. and Molaro, P. and Moortgat, F. and Moskovic, M. and Nagata, N. and Nakamura, K. and Narain, M. and Nason, P. and Nelles, A. and Neubert, M. and Nir, Y. and O'Connell, H. B. and O'Hare, C. A. J. and Olive, K. A. and Peacock, J. A. and Pianori, E. and Pich, A. and Piepke, A. and Pietropaolo, F. and Pomarol, A. and Pordes, S. and Profumo, S. and Quadt, A. and Rabbertz, K. and Rademacker, J. and Raffelt, G. and {Ramsey-Musolf}, M. and Richardson, P. and Ringwald, A. and Robinson, D. J. and Roesler, S. and Rolli, S. and Romaniouk, A. and Rosenberg, L. J and Rosner, J. L. and Rybka, G. and Ryskin, M. G. and Ryutin, R. A. and Safdi, B. and Sakai, Y. and Sarkar, S. and Sauli, F. and Schneider, O. and Sch{\"o}nert, S. and Scholberg, K. and Schwartz, A. J. and Schwiening, J. and Scott, D. and Sefkow, F. and Seljak, U. and Sharma, V. and Sharpe, S. R. and Shiltsev, V. and Signorelli, G. and Silari, M. and Simon, F. and Sj{\"o}strand, T. and Skands, P. and Skwarnicki, T. and Smoot, G. F. and Soffer, A. and Sozzi, M. S. and Spiering, C. and Stahl, A. and Sumino, Y. and Takahashi, F. and Tanabashi, M. and Tanaka, J. and Ta{\v s}evsk{\'y}, M. and Terao, K. and Terashi, K. and Terning, J. and Thoma, U. and Thorne, R. S. and Tiator, L. and Titov, M. and Tovey, D. R. and Trabelsi, K. and Urquijo, P. and Valencia, G. and {Van de Water}, R. and Varelas, N. and Verde, L. and Vivarelli, I. and Vogel, P. and Vogelsang, W. and Vorobyev, V. and Wakely, S. P. and Walkowiak, W. and Walter, C. W. and Wands, D. and Weinberg, D. H. and Weinberg, E. J. and Wermes, N. and White, M. and Wiencke, L. R. and Willocq, S. and Woody, C. L. and Workman, R. L. and Yao, W.-M. and Yokoyama, M. and Yoshida, R. and Zanderighi, G. and Zeller, G. P. and Zhu, R.-Y. and Zhu, S.-L. and Zimmermann, F. and Zyla, P. A. and {Particle Data Group Collaboration} and Anderson, J. and Kramer, M. and Schaffner, P. and Zheng, W.},
  collaboration = {{Particle Data Group}},
  year = {2024},
  journal = {Phys. Rev. D},
  volume = {110},
  number = {3},
  pages = {030001},
  publisher = {American Physical Society},
  doi = {10.1103/PhysRevD.110.030001},
  urldate = {2025-07-21},
}

@article{wilksLargeSampleDistributionLikelihood1938,
  title = {The {{Large-Sample Distribution}} of the {{Likelihood Ratio}} for {{Testing Composite Hypotheses}}},
  author = {Wilks, S. S.},
  year = {1938},
  journal = {The Annals of Mathematical Statistics},
  volume = {9},
  number = {1},
  pages = {60--62},
  publisher = {Institute of Mathematical Statistics},
  issn = {0003-4851, 2168-8990},
  doi = {10.1214/aoms/1177732360},
  urldate = {2023-11-10}
}

@article{t2kcollaborationPreciseMeasurementNeutrino2014,
  title = {Precise {{Measurement}} of the {{Neutrino Mixing Parameter}} $\theta_{23}$ from {{Muon Neutrino Disappearance}} in an {{Off-Axis Beam}}},
  author = {Abe, K. and Adam, J. and Aihara, H. and Akiri, T. and Andreopoulos, C. and Aoki, S. and Ariga, A. and Ariga, T. and Assylbekov, S. and Autiero, D. and Barbi, M. and Barker, G. J. and Barr, G. and Bass, M. and Batkiewicz, M. and Bay, F. and Bentham, S. W. and Berardi, V. and Berger, B. E. and Berkman, S. and Bertram, I. and Bhadra, S. and Blaszczyk, F. d. M. and Blondel, A. and Bojechko, C. and Bordoni, S. and Boyd, S. B. and Brailsford, D. and Bravar, A. and Bronner, C. and Buchanan, N. and Calland, R. G. and Caravaca Rodr{\'i}guez, J. and Cartwright, S. L. and Castillo, R. and Catanesi, M. G. and Cervera, A. and Cherdack, D. and Christodoulou, G. and Clifton, A. and Coleman, J. and Coleman, S. J. and Collazuol, G. and Connolly, K. and Cremonesi, L. and Dabrowska, A. and Danko, I. and Das, R. and Davis, S. and {de Perio}, P. and De Rosa, G. and Dealtry, T. and Dennis, S. R. and Densham, C. and Di Lodovico, F. and Di Luise, S. and Drapier, O. and Duboyski, T. and Duffy, K. and Dufour, F. and Dumarchez, J. and Dytman, S. and Dziewiecki, M. and Emery, S. and Ereditato, A. and Escudero, L. and Finch, A. J. and Floetotto, L. and Friend, M. and Fujii, Y. and Fukuda, Y. and Furmanski, A. P. and Galymov, V. and Giffin, S. and Giganti, C. and Gilje, K. and Goeldi, D. and Golan, T. and Gonin, M. and Grant, N. and Gudin, D. and Hadley, D. R. and Haesler, A. and Haigh, M. D. and Hamilton, P. and Hansen, D. and Hara, T. and Hartz, M. and Hasegawa, T. and Hastings, N. C. and Hayato, Y. and Hearty, C. and Helmer, R. L. and Hierholzer, M. and Hignight, J. and Hillairet, A. and Himmel, A. and Hiraki, T. and Hirota, S. and Holeczek, J. and Horikawa, S. and Huang, K. and Ichikawa, A. K. and Ieki, K. and Ieva, M. and Ikeda, M. and Imber, J. and Insler, J. and Irvine, T. J. and Ishida, T. and Ishii, T. and Ives, S. J. and Iwai, E. and Iyogi, K. and Izmaylov, A. and Jacob, A. and Jamieson, B. and Johnson, R. A. and Jo, J. H. and Jonsson, P. and Jung, C. K. and Kabirnezhad, M. and Kaboth, A. C. and Kajita, T. and Kakuno, H. and Kameda, J. and Kanazawa, Y. and Karlen, D. and Karpikov, I. and Kearns, E. and Khabibullin, M. and Khotjantsev, A. and Kielczewska, D. and Kikawa, T. and Kilinski, A. and Kim, J. and Kisiel, J. and Kitching, P. and Kobayashi, T. and Koch, L. and Kolaceke, A. and Konaka, A. and Kormos, L. L. and Korzenev, A. and Koseki, K. and Koshio, Y. and Kreslo, I. and Kropp, W. and Kubo, H. and Kudenko, Y. and Kumaratunga, S. and Kurjata, R. and Kutter, T. and Lagoda, J. and Laihem, K. and Lamont, I. and Laveder, M. and Lawe, M. and Lazos, M. and Lee, K. P. and Lindner, T. and Lister, C. and Litchfield, R. P. and Longhin, A. and Ludovici, L. and Macaire, M. and Magaletti, L. and Mahn, K. and Malek, M. and Manly, S. and Marino, A. D. and Marteau, J. and Martin, J. F. and Maruyama, T. and Marzec, J. and Mathie, E. L. and Matveev, V. and Mavrokoridis, K. and Mazzucato, E. and McCarthy, M. and McCauley, N. and McFarland, K. S. and McGrew, C. and Metelko, C. and Mezzetto, M. and Mijakowski, P. and Miller, C. A. and Minamino, A. and Mineev, O. and Mine, S. and Missert, A. and Miura, M. and Monfregola, L. and Moriyama, S. and Mueller, {\relax Th}. A. and Murakami, A. and Murdoch, M. and Murphy, S. and Myslik, J. and Nagasaki, T. and Nakadaira, T. and Nakahata, M. and Nakai, T. and Nakamura, K. and Nakayama, S. and Nakaya, T. and Nakayoshi, K. and Naples, D. and Nielsen, C. and Nirkko, M. and Nishikawa, K. and Nishimura, Y. and O'Keeffe, H. M. and Ohta, R. and Okumura, K. and Okusawa, T. and Oryszczak, W. and Oser, S. M. and Owen, R. A. and Oyama, Y. and Palladino, V. and Palomino, J. and Paolone, V. and Payne, D. and Perevozchikov, O. and Perkin, J. D. and Petrov, Y. and Pickard, L. and Pinzon Guerra, E. S. and Pistillo, C. and Plonski, P. and Poplawska, E. and Popov, B. and Posiadala, M. and Poutissou, J.-M. and Poutissou, R. and Przewlocki, P. and Quilain, B. and Radicioni, E. and Ratoff, P. N. and Ravonel, M. and Rayner, M. A. M. and Redij, A. and Reeves, M. and {Reinherz-Aronis}, E. and Retiere, F. and Robert, A. and Rodrigues, P. A. and Rojas, P. and Rondio, E. and Roth, S. and Rubbia, A. and Ruterbories, D. and Sacco, R. and Sakashita, K. and S{\'a}nchez, F. and Sato, F. and Scantamburlo, E. and Scholberg, K. and Schoppmann, S. and Schwehr, J. and Scott, M. and Seiya, Y. and Sekiguchi, T. and Sekiya, H. and Sgalaberna, D. and Shiozawa, M. and Short, S. and Shustrov, Y. and Sinclair, P. and Smith, B. and Smith, R. J. and Smy, M. and Sobczyk, J. T. and Sobel, H. and Sorel, M. and Southwell, L. and Stamoulis, P. and Steinmann, J. and Still, B. and Suda, Y. and Suzuki, A. and Suzuki, K. and Suzuki, S. Y. and Suzuki, Y. and Szeglowski, T. and Tacik, R. and Tada, M. and Takahashi, S. and Takeda, A. and Takeuchi, Y. and Tanaka, H. K. and Tanaka, H. A. and Tanaka, M. M. and Terhorst, D. and Terri, R. and Thompson, L. F. and Thorley, A. and Tobayama, S. and Toki, W. and Tomura, T. and Totsuka, Y. and Touramanis, C. and Tsukamoto, T. and Tzanov, M. and Uchida, Y. and Ueno, K. and Vacheret, A. and Vagins, M. and Vasseur, G. and Wachala, T. and Waldron, A. V. and Walter, C. W. and Wark, D. and Wascko, M. O. and Weber, A. and Wendell, R. and Wilkes, R. J. and Wilking, M. J. and Wilkinson, C. and Williamson, Z. and Wilson, J. R. and Wilson, R. J. and Wongjirad, T. and Yamada, Y. and Yamamoto, K. and Yanagisawa, C. and Yen, S. and Yershov, N. and Yokoyama, M. and Yuan, T. and Yu, M. and Zalewska, A. and Zalipska, J. and Zambelli, L. and Zaremba, K. and Ziembicki, M. and Zimmerman, E. D. and Zito, M. and {\.Z}muda, J.},
  collaboration={T2K},
  year = 2014,
  journal = {Phys. Rev. Lett.},
  volume = {112},
  number = {18},
  pages = {181801},
  publisher = {American Physical Society},
  doi = {10.1103/PhysRevLett.112.181801},
  urldate = {2025-11-10},
}

@article{feldmanUnifiedApproachClassical1998,
  title = {Unified Approach to the Classical Statistical Analysis of Small Signals},
  author = {Feldman, Gary J. and Cousins, Robert D.},
  year = 1998,
  journal = {Phys. Rev. D},
  volume = {57},
  number = {7},
  pages = {3873--3889},
  publisher = {American Physical Society},
  doi = {10.1103/PhysRevD.57.3873},
  urldate = {2022-12-20},
}

\end{document}